\documentclass{article}[12pt]

\usepackage[doublespacing]{setspace}
\usepackage[utf8]{inputenc}
\usepackage[top=1in, bottom=1in, lmargin=1in, rmargin=1in]{geometry}
\usepackage[authoryear,semicolon]{natbib}
\usepackage{authblk}
\usepackage[dvipsnames]{xcolor}
\usepackage{parskip, amsmath, amssymb, mathtools, bm, comment}
\usepackage{graphicx}
\usepackage{algorithm, algpseudocode}
\usepackage{scalerel} 

\usepackage{clipboard}

    \usepackage[hypertexnames=false]{hyperref}
    
\providecommand{\keywords}[1]{\textbf{\textit{Keywords: }} #1} 

\makeatletter
\def\thanks#1{\protected@xdef\@thanks{\@thanks
        \protect\footnotetext{#1}}}
\makeatother

\newcommand{\1}{\mathbf{1}}
\renewcommand{\d}{\, \mathrm{d}}
\newcommand{\A}{\mathbf{A}}
\newcommand{\B}{\mathbf{B}}

\newcommand{\G}{\mathbf{G}}
\renewcommand{\H}{\mathbf{H}}
\newcommand{\I}{\mathbf{I}}
\newcommand{\J}{\mathbf{J}}

\renewcommand{\L}{\mathbf{L}}
\newcommand{\M}{\mathbf{M}}

\newcommand{\W}{\mathbf{W}}
\newcommand{\X}{\mathbf{X}}

\newcommand{\Z}{\mathbf{Z}}

\renewcommand{\u}{\mathbf{u}}

\newcommand{\y}{\mathbf{y}}

\newcommand{\x}{\mathbf{x}}
\newcommand{\z}{\mathbf{z}}

\newcommand{\0}{\mathbf{0}}

\newcommand{\balpha}{\bm{\alpha}}
\newcommand{\bbeta}{\bm{\beta}}
\newcommand{\bgamma}{\bm{\gamma}}
\newcommand{\bzeta}{\bm{\zeta}}
\newcommand{\blambda}{\bm{\lambda}}
\newcommand{\btheta}{\bm{\theta}}

\newcommand{\btau}{\bm{\tau}}

\newcommand{\bepsilon}{\bm{\varepsilon}}
\newcommand{\bmu}{\bm{\mu}}

\newcommand{\bSigma}{\mathbf{\Sigma}}

\newcommand{\Gau}{\textrm{Gau}}

\newcommand{\Exp}{\mathbb{E}}

\newcommand{\Tr}{\operatorname{Tr}}

\newcommand{\logit}{\text{logit}}
\newcommand{\ind}{\perp\!\!\!\!\perp} 
\DeclareMathOperator*{\argmin}{arg\,min}

\newcommand\wh[1]{\hstretch{4}{\widehat{\hstretch{.25}{#1}}}} 
\newcommand\wtilde[1]{\hstretch{4}{\tilde{\hstretch{.25}{#1}}}}

\DeclarePairedDelimiter\abs{\lvert}{\rvert}%
\DeclarePairedDelimiter\norm{\lVert}{\rVert}%

\DeclarePairedDelimiterX{\divergence}[2]{(}{)}{%
  #1\;\delimsize\|\;#2%
}

\makeatletter
\let\oldabs\abs
\def\abs{\@ifstar{\oldabs}{\oldabs*}}
\let\oldnorm\norm
\def\norm{\@ifstar{\oldnorm}{\oldnorm*}}
\makeatother

\title{R-VGAL: A Sequential Variational Bayes Algorithm for Generalised Linear Mixed Models}

\author[1, 2, *]{Bao Anh Vu}
\author[1]{David Gunawan}
\author[1, 2]{Andrew Zammit-Mangion\thanks{* Corresponding author \\ E-mail address: bavu@uow.edu.au}}

\affil[1]{School of Mathematics and Statistics, University of Wollongong, Wollongong, New South Wales, Australia}
\affil[2] {Securing Antarctica’s Environmental Future, University of Wollongong, Wollongong, New South Wales, Australia}

\date{\today}

\begin{document}

\maketitle

\begin{abstract}
    Models with random effects, such as generalised linear mixed models (GLMMs), are often used for analysing clustered data. Parameter inference with these models is difficult because of the presence of cluster-specific random effects, which must be integrated out when evaluating the likelihood function. Here, we propose a sequential variational Bayes algorithm, called Recursive Variational Gaussian Approximation for Latent variable models (R-VGAL), for estimating parameters in GLMMs. The R-VGAL algorithm operates on the data sequentially, requires only a single pass through the data, and can provide parameter updates as new data are collected without the need of re-processing the previous data. At each update, the R-VGAL algorithm requires the gradient and Hessian of a ``partial" log-likelihood function evaluated at the new observation, which are generally not available in closed form for GLMMs. To circumvent this issue, we propose using an importance-sampling-based approach for estimating the gradient and Hessian via Fisher's and Louis' identities. We find that R-VGAL can be unstable when traversing the first few data points, but that this issue can be mitigated by introducing a damping factor in the initial steps of the algorithm. Through illustrations on both simulated and real datasets, we show that R-VGAL provides good approximations to posterior distributions, that it can be made robust through damping, and that it is computationally efficient.
    
\end{abstract}

\keywords{Fisher's identity, intractable gradient, latent variable model, Louis' identity,
damped Newton's method} 
\section{Introduction}

Mixed models are useful for analysing clustered data, wherein observations that come from the same cluster/group are likely to be correlated. Example datasets include records of students clustered within schools, and repeated measurements of biomarkers on patients. Mixed models account for intra-group dependencies by incorporating cluster/group-specific ``random effects". Inference with these models is made challenging by the fact that the likelihood function involves integrals over the random effects that are not usually tractable except for the few cases where the distribution of the random effects is conjugate to the distribution of the data, such as in the linear mixed model~\citep{verbeke1997linear}, the beta-binomial model~\citep{crowder1979inference}, and Rasch's Poisson count model~\citep{jansen1994parameters}. Notably, there is no closed-form expression for the likelihood function in the case of the ubiquitous logistic mixed model.  
    
Maximum-likelihood-based approaches are often used for parameter inference in mixed models. In the case of linear mixed models, parameter inference via maximum likelihood estimation is straightforward~\citep[e.g.,][]{wakefield2013bayesian}. For mixed models with an intractable likelihood, integrals over random effects need to be numerically approximated, for example by using Gaussian quadrature~\citep{naylor1982applications} or the Laplace approximation~\citep{tierney1986accurate}. The likelihood may also be indirectly maximised using an expectation-maximisation type algorithm~\citep{dempster1977maximum}, which treats the random effects as missing, and iteratively maximises the ``expected complete-data log-likelihood" of the data and the random effects. Quasi-likelihood approaches such as penalised quasi-likelihood~\citep[PQL,][]{breslow1993approximate} and marginal quasi-likelihood~\citep[MQL,][]{goldstein1991nonlinear} approximate nonlinear mixed models with linear mixed models, so that well-developed estimation routines for linear mixed models can be applied; see~\cite{tuerlinckx2006statistical} for a detailed discussion of these methods. These maximum-likelihood-based methods provide point estimates and not full posterior distributions over the parameters.
    
Full posterior distributions can be obtained using Markov chain Monte Carlo~\citep[MCMC, e.g.,][]{zhao2006general, fong2010bayesian}. MCMC provides exact, sample-based posterior distributions, but at a higher computational cost than maximum-likelihood-based methods. Alternatively, variational Bayes (VB) methods \citep[e.g.,][]{ong2018gaussian, tan2018gaussian} are becoming increasingly popular for estimating parameters in complex statistical models. These methods approximate the exact posterior distribution with a member from a simple and tractable family of distributions; this family is usually chosen to balance the accuracy of the approximation against the computational cost required to obtain the approximation. VB methods are usually computationally cheaper than MCMC methods. VB approaches can either 
batch-process the data~\citep[e.g.,][]{tran2016parallel, ong2018gaussian, tan2018gaussian} or sequentially process data points~\citep[e.g.,][]{broderick2013streaming, gunawan2021variational, lambert2022recursive}.
For settings with large amounts of data, a method that targets the posterior distribution via sequential processing of the data offers several advantages. 
The so-called Recursive Variational Gaussian Approximation \citep[R-VGA,][]{lambert2022recursive} algorithm is a recently-developed sequential variational Bayes method that provides a fast and accurate approximation to the posterior distribution with only one pass through the data, making it computationally efficient when compared to MCMC or batch variational Bayes. \citet{lambert2022recursive} apply the R-VGA algorithm to linear and logistic regression models without random effects. 

In this paper, we build on the R-VGA algorithm by proposing a novel recursive variational Gaussian approximation, called Recursive Variational Gaussian Approximation for Latent variable models (R-VGAL), for estimating the parameters in GLMMs. At each update, R-VGAL requires the gradient and Hessian of the ``partial" log-likelihood evaluated at the new observation, which are often not available in closed form. To circumvent this issue, we propose an importance-sampling-based approach for estimating the gradient and Hessian that uses Fisher's and Louis' identities ~\citep{cappe2005}. This approach was inspired by the work of ~\cite{nemeth2016particle}, who used Fisher's and Louis' identities to approximate the gradient and Hessian in a sequential Monte Carlo context. The efficacy of R-VGAL is illustrated using linear, logistic and Poisson mixed effect models on simulated and real datasets. The examples show that R-VGAL provides good approximations to the exact posterior distributions estimated using Hamiltonian Monte Carlo~\citep[HMC,][]{neal2011, betancourt2015hamiltonian} and at a low computational cost.

The paper is organised as follows. Sect.~\ref{sec:methodology} provides some background on the sequential variational Bayes framework and presents the R-VGAL algorithm. Sect.~\ref{sec:applications} applies the R-VGAL algorithm to simulated and real datasets. Sect.~\ref{sec:conclusion} concludes with a discussion of our results and an overview of future research directions. This article has an online supplement containing additional technical details, and the code to reproduce results from the simulation and real-data experiments is available on \url{https://github.com/bao-anh-vu/R-VGAL}.

\section{The R-VGAL algorithm}
\label{sec:methodology}

This section reviews GLMMs~\citep[e.g.][]{demidenko2013mixed, faraway2016extending} and provides some background on the R-VGA algorithm of~\cite{lambert2022recursive}, and then introduces the R-VGAL algorithm for making parameter inference with GLMMs. 

\subsection{Generalised linear mixed models}
\label{sec:glmms}

GLMMs are statistical models that contain both fixed effects and random effects. Typically, the fixed effects are common across groups, while the random effects are group-specific, and this is the setting we focus on. We briefly discuss the potential application of R-VGAL to models with more complicated random effect structures, such as crossed or nested random effects, in Sect.~\ref{sec:crossed_nested} of the online supplement.

Denote by $y_{ij}$ the $j$th response in the $i$th group, for $i = 1, \dots, N$ groups and $j = 1, \dots, n_i$, where $n_i$ is the number of responses in group $i$. Let $\y \equiv ({\y_1^\top,\dots,\y_N^\top})^\top$ be a vector of observations, where ${\y_i \equiv (y_{i1}, \dots, y_{in_i})^\top}$ are the responses from the $i$th group. The GLMMs we consider are constructed by first assigning each $y_{ij}$ a distribution $y_{ij} \mid \bbeta, \balpha_i, \phi \sim p(\cdot)$, where $p(\cdot)$ is a member of the exponential family with a dispersion parameter $\phi$ that is usually related to the variance of the datum, $\bbeta$ are the fixed effect parameters, and $\balpha_i$ are the group-specific random effects for $i = 1, \dots, N$.
Then, the mean of the responses, $\mu_{ij} \equiv \Exp(y_{ij} \mid \bbeta, \balpha_i, \phi)$, is modelled as
\begin{equation}
    g(\mu_{ij}) = \x_{ij}^\top \bbeta + \z_{ij}^\top \balpha_i, \quad i = 1, \dots, N, \quad j = 1, \dots, n_i, 
\end{equation}
where $\x_{ij}$ is a vector of fixed effect covariates corresponding to the $j$th response in the $i$th group;
$\z_{ij}$ is a vector of predictor variables corresponding to the $j$th response and the $i$th random effect; and $g(\cdot)$ is a {link function} that links the response mean $\mu_{ij}$ to the linear predictor $\x_{ij}^\top \bbeta + \z_{ij}^\top \balpha_i$. We further assume that $\balpha_i \ind \balpha_{i'}$ for $i \neq i'$. The random effects $\balpha_i$, for $i = 1, \dots, N$, are assumed to follow a normal distribution with mean $\0$ and covariance matrix $\bSigma_\alpha$, that is, each $\balpha_i \mid \bSigma_\alpha \sim \Gau(\0, \bSigma_\alpha)$. In practice, some structure is often assumed for the random effects covariance matrix so that it is parameterised in terms of a smaller number of parameters $\btau$, that is,  $\bSigma_\alpha = \bSigma_\alpha (\btau)$. Inference is then made on the parameters $\btheta = (\bbeta^\top, \btau^\top, \phi)^\top$. 

The main objective of Bayesian inference is to obtain the posterior distribution of the model parameters $\btheta$ given the observations $\y$ and the prior distribution $p(\btheta)$. Through Bayes' rule, the posterior distribution of $\btheta$ is 
\begin{equation}
    p(\btheta \mid \y) = p(\bbeta, \btau, \phi \mid \y) \propto p(\y \mid \bbeta, \btau, \phi) p (\bbeta, \btau, \phi).
\end{equation}
The likelihood function, 
\begin{equation}
\label{eq:likelihood_integral}
    p(\y \mid \bbeta, \btau, \phi) = \prod_{i=1}^N \int p(\y_i \mid \balpha_i, \bbeta, \phi) p(\balpha_i \mid \btau) \d \balpha_i,
\end{equation}
involves integrals over the random effects $\balpha_i, i = 1, \dots, N$. The likelihood function can be calculated exactly for the linear mixed model with normally distributed random effects, for which
\begin{equation}
    y_{ij} = \x_{ij}^\top \bbeta + \z_{ij}^\top \balpha_i + \epsilon_{ij}, \quad \balpha_i \sim \Gau(\0, \bSigma_\alpha(\btau)), \quad \epsilon_{ij} \sim \Gau(0, \sigma_\epsilon^2),
\end{equation}
for $i = 1, \dots, N$ and $j = 1, \dots, n_i$, where $\epsilon_{ij}$ is a zero-mean, independent, normally distributed error term with variance $\sigma_\epsilon^2$ that is associated with the $j$th response from the $i$th group. At the group level, this model can be written as 
\begin{equation*}
    \y_i = \X_i \bbeta + \Z_i \balpha_i + \bepsilon_i, \quad \balpha_i \sim \Gau(\0, \bSigma_\alpha(\btau)), \quad \bepsilon_i \sim \Gau(\0, \sigma_\epsilon^2 \I_{n_i}),
\end{equation*}
where $\X_i \equiv (\x_{i1}, \dots, \x_{in_i})^\top$, $\Z_i \equiv (\z_{i1}, \dots, \z_{in_i})^\top$, and $\bepsilon_i \equiv (\epsilon_{i1}, \dots, \epsilon_{in_i})^\top$, with $n_i$ being the number of observations in the $i$th group, for $i = 1, \dots, N$, and $\I_m$ denotes an identity matrix of size $m \times m$. The likelihood function for this linear mixed model is
\begin{equation}
    p(\y \mid \bbeta, \btau, \sigma_\epsilon^2) = \prod_{i = 1}^{N} p(\y_{i} \mid \bbeta, \btau, \sigma_\epsilon^2) = \prod_{i = 1}^{N} \Gau(\X_i \bbeta, \Z_i \bSigma_\alpha(\btau) \Z_i^\top + \sigma_\epsilon^2 \I_{n_i}).
\end{equation}
The gradient and Hessian of the log-likelihood for the linear mixed model are also available in closed form. However, the likelihood $p(\y_i \mid \balpha_i, \bbeta, \phi)$ in~\eqref{eq:likelihood_integral} cannot be computed exactly for general random effects models. One important case is the logistic mixed model given by
\begin{equation}
    y_{ij} \sim \text{Bernoulli}(\pi_{ij}), \quad \logit(\pi_{ij}) = \x_{ij}^\top \bbeta + \z_{ij}^\top \balpha_i, \quad i = 1, \dots, N, \quad j = 1, \dots, n_i,
\end{equation}
where $\logit(\pi_{ij}) = \log \left(\frac{\pi_{ij}}{1 - \pi_{ij}} \right)$. The gradient and Hessian of the log-likelihood function for this model can, however, be estimated unbiasedly, as we show in Sects.~\ref{sec:fishers_identity} and~\ref{sec:louis_identity}.

\subsection{Sequential VB and R-VGA}
\label{sec:rvga_and_rvgal}

We begin this section with a review of VB and the sequential VB framework. We then present the main steps in the derivations of the R-VGA algorithm of \cite{lambert2022recursive}, on which our algorithm is based.

\subsubsection{Sequential VB}
\label{sec:sequential_vb}
VB is usually used for posterior inference in complex statistical models when inference using asymptotically exact methods such as MCMC is too costly;
for a review see, for example, \citet{blei2017variational}.
Let $\boldsymbol{\theta}$ be a vector of model parameters. Here, we consider the class of VB methods where
the posterior distribution $p(\btheta \mid \y)$ is approximated by a tractable density $q(\btheta; \blambda)$ parameterised by $\blambda$. The variational parameters $\blambda$ are optimised by 
minimising the Kullback-Leibler (KL) divergence between the variational distribution and the posterior distribution, that is, by minimising
\begin{equation}
    \text{KL} \divergence{q(\btheta; \blambda)}{p(\btheta \mid \y)} \equiv \int q(\btheta; \blambda) \log \frac{q(\btheta; \blambda)}{p(\btheta \mid \y)} \d \btheta.
\end{equation}
Many VB algorithms require processing the data as a batch; see, for example, \cite{ong2018gaussian} and~\cite{tan2018gaussian}. The variational parameters $\blambda$ are typically updated in an iterative manner using stochastic gradient descent~\citep[SGD,][]{Hoffman:2013, Kingma2014}. 
In settings with large amounts of data or continuously-arriving data, it is often more practical to use online or sequential variational Bayes algorithms that update the approximation to the posterior distribution sequentially as new observations become available. These online/sequential algorithms are designed to handle data that are too large to fit in memory or that arrive in a continuous stream.

In a sequential VB framework, such as that proposed by \cite{broderick2013streaming}, the observations $\y_1, \dots, \y_N$ are incorporated sequentially so that at iteration $i$, $i = 1, \dots, N$, one targets an approximation $q_i(\btheta) \equiv q(\btheta; \blambda_i)$ that is closest in a KL sense to 
the ``pseudo-posterior" $p(\y_i \mid \btheta) q_{i-1}(\btheta)/\mathcal{Z}_i$, where
\begin{equation}
    \mathcal{Z}_i \equiv \int p(\y_i \mid \btheta) q_{i-1}(\btheta) \d \btheta.
\end{equation}
In this framework, $q_{i-1}(\btheta)$ is treated as the ``prior" for the next iteration $i$, and the KL divergence between $q_i(\btheta)$ and the ``pseudo-posterior" 
is minimised at each iteration.  \cite{broderick2013streaming} use a mean field VB approach~\citep[e.g.,][]{ormerod2010explaining}, which assumes no posterior dependence between the elements of $\btheta$. The R-VGA algorithm proposed by~\cite{lambert2022recursive} follows closely that of~\cite{broderick2013streaming}, but uses a variational distribution of the form $q_i(\btheta) = \Gau(\bmu_i, \bSigma_i)$, where $\bSigma_i$ is a full covariance matrix, and seeks closed-form updates for $\blambda_i \equiv \{\bmu_i,\bSigma_i\}$ that minimise the KL divergence between $q_i(\btheta)$ and ${p(\y_i \mid \btheta) q_{i-1}(\btheta) / \mathcal{Z}_i}$ for $i = 1, \dots, N$. Another sequential VB algorithm that is similar to that of~\cite{broderick2013streaming} is the Updating Variational Bayes~\citep[UVB,][]{tomasetti2022updating} algorithm, which uses SGD~\citep{bottou2010large} at every iteration, $i = 1, \dots, N$, to minimise the KL divergence between $q_i(\btheta)$ and ${p(\y_i \mid \btheta) q_{i-1}(\btheta)/\mathcal{Z}_i}$. One advantage of UVB compared to R-VGA is that it does not have to assume that the prior and variational distributions are Gaussian; see Sect. 5.2 of~\cite{tomasetti2022updating} for an example of UVB where a beta prior is used for one of the parameters and the variational distribution is a mixture of multivariate normal distributions. However, due to the lack of restrictions on the form of the variational distribution, UVB requires running a full optimisation algorithm at each iteration, whereas the R-VGAL updates are available in closed form. 

Detailed derivations for the R-VGA algorithm can be found in~\cite{lambert2022recursive}. We provide below a sketch of the derivations to aid the exposition of the methodology in subsequent sections.

\subsubsection{The R-VGA algorithm}
\label{sec:rvga}
Denote by $\y_{1:i} \equiv (\y_1^\top, \dots, \y_i^\top)^\top$ a collection of observations from groups $1$ to $i$, $i = 1, \dots, N$. By assumption of conditional independence between observations $\y_1, \dots, \y_i$ given the parameters $\btheta$, the KL divergence between the variational distribution $q_i(\btheta)$ and the posterior distribution $p(\btheta \mid \y_{1:i})$ can be expressed as
\begin{align*}
    \text{KL} \divergence{q_i(\btheta)}{p(\btheta \mid \y_{1:i})} 
    &\equiv \int q_i(\btheta) \log \frac{q_i(\btheta)}{p(\btheta \mid \y_{1:i})} \d \btheta \\
    &= \Exp_{q_i} \left(\log q_i(\btheta) - \log p(\btheta \mid \y_{1:i-1}) - \log p(\y_i \mid \btheta) \right) + \log p(\y_{1:i}) - \log p(\y_{1:i-1}).
\end{align*}
The posterior distribution after incorporating the first $i-1$ groups of observations, $p(\btheta \mid \y_{1:i-1})$, is approximated by the variational distribution $q_{i-1} (\btheta)$ to give
\begin{equation}
    \label{eq:KL_approx}
    \text{KL} \divergence{q_i(\btheta)}{p(\btheta \mid \y_{1:i})} \approx \Exp_{q_i}(\log q_i(\btheta) - \log q_{i-1} (\btheta) - \log p(\y_i \mid \btheta)) + \log p(\y_{1:i}) - \log p(\y_{1:i-1}).
\end{equation}
The R-VGA algorithm assumes a variational distribution of the form $q_i(\btheta) = \Gau(\bmu_i, \bSigma_i)$ and seeks parameters $\bmu_i$ and $\bSigma_i$ that minimise~\eqref{eq:KL_approx}. As the last two terms in the right hand side of~\eqref{eq:KL_approx} do not depend on $\btheta$, the optimisation problem is equivalent to finding 
\begin{equation}
    \label{eq:exp_to_minimise}
    \argmin_{\bmu_i, \bSigma_i} \, \Exp_{q_i}(\log q_i(\btheta) - \log q_{i-1} (\btheta) - \log p(\y_i \mid \btheta)).
\end{equation}
Differentiating the expectation~\eqref{eq:exp_to_minimise} with respect to $\bmu_i$ and $\bSigma_i$, setting the derivatives to zero, and rearranging the resulting equations, yields the following recursive updates for the variational mean~$\bmu_i$ and precision matrix~$\bSigma_i^{-1}$:
\begin{align}
    \bmu_i &= \bmu_{i-1} + \bSigma_{i-1} \nabla_{\bmu_i} \Exp_{q_i}(\log p(\y_i \mid \btheta)), \label{eq:bmu_1}\\
    \bSigma_i^{-1} &= \bSigma_{i-1}^{-1} - 2\nabla_{\bSigma_i} \Exp_{q_i}(\log p(\y_i \mid \btheta)). \label{eq:bSigma_1}
\end{align}
Then, using Bonnet's Theorem~\citep{bonnet1964transformations} on~\eqref{eq:bmu_1} and Price's Theorem~\citep{price1958useful} on~\eqref{eq:bSigma_1}, we rewrite the gradient terms as
\begin{align}
    \nabla_{\bmu_i} \Exp_{q_i}(\log p(\y_i \mid \btheta)) &= \Exp_{q_i}(\nabla_{\btheta} \log p(\y_i \mid \btheta)), \\
    \nabla_{\bSigma_i} \Exp_{q_i}(\log p(\y_i \mid \btheta)) &= \frac{1}{2} \Exp_{q_i}(\nabla_{\btheta}^2 \log p(\y_i \mid \btheta)).
\end{align}
Thus the updates~\eqref{eq:bmu_1} and~\eqref{eq:bSigma_1} become
\begin{align}
    \bmu_i &= \bmu_{i-1} + \bSigma_{i-1} \Exp_{q_i}(\nabla_{\btheta} \log p(\y_i \mid \btheta)), \label{eq:bmu_2}\\
    \bSigma_i^{-1} &= \bSigma_{i-1}^{-1} - \Exp_{q_i}(\nabla_{\btheta}^2 \log p(\y_i \mid \btheta)). \label{eq:bSigma_2}
\end{align}
These updates are implicit as they require the evaluation of expectations with respect to $q_i(\btheta)$. Under the assumption that $q_i(\btheta)$ is close to $q_{i-1}(\btheta)$, \cite{lambert2022recursive} propose replacing $q_i(\btheta)$ with $q_{i-1}(\btheta)$ in~\eqref{eq:bmu_2} and~\eqref{eq:bSigma_2}, and replacing $\bSigma_{i-1}$ with $\bSigma_i$ on the right hand side of~\eqref{eq:bmu_2}, to yield an explicit scheme
\begin{align}
    \bmu_i &= \bmu_{i-1} + \bSigma_i \Exp_{q_{i-1}}(\nabla_{\btheta} \log p(\y_i \mid \btheta)), \label{eq:bmu_3}\\
    \bSigma_i^{-1} &= \bSigma_{i-1}^{-1} - \Exp_{q_{i-1}}(\nabla_{\btheta}^2 \log p(\y_i \mid \btheta)). \label{eq:bSigma_3}
\end{align}
Equations~\eqref{eq:bmu_3} and~\eqref{eq:bSigma_3} form the so-called R-VGA algorithm of~\cite{lambert2022recursive}.

\Copy{order_one}{We note that an ``order 1 form" of the R-VGA algorithm exists, which allows the variational precision matrix to be updated using the first order derivatives of the log-likelihood without the need for the Hessian matrix. However, these updates are implicit and not directly implementable. Corollary 1 of~\cite{lambert2022recursive} provides more details on this Hessian-free form.} 

\subsection{R-VGAL}
\label{sec:rvgal}
The R-VGA updates in~\eqref{eq:bmu_3} and~\eqref{eq:bSigma_3} require the gradient $\nabla_{\btheta} \log p(\y_i \mid \btheta)$ and Hessian ${\nabla_{\btheta}^2 \log p(\y_i \mid \btheta)}$ of the ``partial" log-likelihood for the $i$th observation. However, for the GLMMs discussed in Sect.~\ref{sec:glmms}, there are usually no closed-form expressions for said quantities, as evaluation of the partial log-likelihood involves an intractable integral over the random effects $\balpha_i$. Our R-VGAL algorithm circumvents this issue by replacing $\nabla_{\btheta} \log p(\y_i \mid \btheta)$ and $\nabla^2_{\btheta} \log p(\y_i \mid \btheta)$ with their unbiased estimates, $\wh{\nabla_{\btheta} \log p(\y_i \mid \btheta)}$ and $\wh{\nabla^2_{\btheta} \log p(\y_i \mid \btheta)}$, respectively. 
These unbiased estimates are obtained by using an importance-sampling-based approach applied to Fisher's and Louis' identities~\citep{cappe2005}, which we discuss in more detail in Sects.~\ref{sec:fishers_identity} and~\ref{sec:louis_identity}. We summarise the R-VGAL algorithm in Algorithm~\ref{algo:rvgal}.

To approximate the expectations with respect to $q_{i-1}(\btheta)$ in the updates of the variational mean and precision matrix in Algorithm~\ref{algo:rvgal}, we generate Monte Carlo samples, ${\btheta^{(l)} \sim q_{i-1}(\btheta)}$, $l = 1, \dots, S$, and compute:
\begin{align*}
    \Exp_{q_{i-1}} (\wh{\nabla_{\btheta} \log p(\y_i \mid \btheta)}) &\approx \frac{1}{S} \sum_{l=1}^S \wh{\nabla_{\btheta} \log p(\y_i \mid \btheta^{(l)})}, \\
    \Exp_{q_{i-1}} (\wh{\nabla_{\btheta}^2 \log p(\y_i \mid \btheta)}) &\approx \frac{1}{S} \sum_{l=1}^S \wh{\nabla_{\btheta}^2 \log p(\y_i \mid \btheta^{(l)})},
\end{align*}
for $i = 1, \dots, N$.

\begin{algorithm} [t!]
\caption{R-VGAL} 
\label{algo:rvgal}
\begin{algorithmic}
    \State Input: observations $\y_1, \dots, \y_N$, initial values $\bmu_0$ and $\bSigma_0$.
    \State Output: variational parameters $\bmu_i$ and $\bSigma_i$, for $i=1,...,N$.
    \State Set $q_0(\btheta) = \Gau(\bmu_0, \bSigma_0)$.
    \For {$i = 1, \dots, N$}
    \State $\bmu_i = \bmu_{i-1} + \bSigma_i \Exp_{q_{i-1}} (\wh{\nabla_{\btheta} \log p(\y_i \mid \btheta)})$ 
    \State $\bSigma_i^{-1} = \bSigma_{i-1}^{-1} - \Exp_{q_{i-1}} (\wh{\nabla_{\btheta}^2 \log p(\y_i \mid \btheta)})$ 
    \EndFor
\end{algorithmic}
\end{algorithm}

The following sections discuss approaches to obtain unbiased estimates of the gradient and the Hessian of the log-likelihood with respect to the parameters.

\subsubsection{Approximation of the gradient with Fisher's identity}
\label{sec:fishers_identity}

Consider the $i$th iteration. Fisher's identity~\citep{cappe2005} for the gradient of $\log p(\y_{i} \mid \btheta)$ is
\begin{equation}
\label{eq:fishers_identity}
    \nabla_{\btheta} \log p(\y_i \mid \btheta) = \int p(\balpha_i \mid \y_i, \btheta) \nabla_{\btheta} \log p(\y_i, \balpha_i \mid \btheta) \d \balpha_i.
\end{equation}
If it is possible to sample directly from $p(\balpha_i \mid \y_i, \btheta)$ (e.g., as it is with the linear random effects model in Sect.~\ref{sec:lmm}), the above identity can be approximated by 
\begin{equation}
    \nabla_{\btheta} \log p(\y_i \mid \btheta) \approx \frac{1}{S_\alpha} \sum_{s=1}^{S_\alpha}  \nabla_{\btheta} \log p(\y_i, \balpha_i^{(s)} \mid \btheta), \quad \balpha_i^{(s)} \sim p(\balpha_i \mid \y_i, \btheta).
\end{equation}
In the case where direct sampling from $p(\balpha_i \mid \y_i, \btheta)$ is difficult, we use importance sampling~\citep[e.g.,][]{tokdar2010importance} to estimate the gradient of the log-likelihood in~\eqref{eq:fishers_identity}. 
Specifically, we draw samples ${\{\balpha_i^{(s)}: s = 1, \dots, S_\alpha\}}$ from an importance distribution $r(\balpha_i \mid \y_i, \btheta)$, and then compute the weights
\begin{equation*}
    w^{(s)}_i = \frac{p(\y_i \mid \balpha_i^{(s)}, \btheta) p(\balpha_i^{(s)} \mid \btheta)}{r(\balpha_i^{(s)} \mid \y_i, \btheta)}, \quad s = 1, \dots, S_\alpha.
\end{equation*}
The gradient of the log-likelihood is then approximated as
\begin{equation}
    \nabla_{\btheta} \log p(\y_i \mid \btheta) \approx \sum_{s=1}^{S_\alpha} \bar{w}_i^{(s)} \nabla_{\btheta} \log p(\y_i, \balpha_i^{(s)} \mid \btheta) ,
\end{equation}
where $\mathcal{W}_i \equiv \{\bar{w}^{(s)}_i : s = 1, \dots, S_\alpha\}$ are the normalised weights given by
\begin{equation*}
    \bar{w}^{(s)}_i = \frac{w^{(s)}_i}{\sum_{q=1}^{S_\alpha} w^{(q)}_i}, \quad s = 1, \dots, S_\alpha.
\end{equation*}
One possible choice for the importance distribution is the distribution of the random effects, that is, $p(\balpha_i \mid \btheta)$. In this case, the weights $\mathcal{W}_i$ reduce to
\begin{equation*}
    w^{(s)}_i = p(\y_i \mid \balpha_i^{(s)}, \btheta), \quad s = 1, \dots, S_\alpha.
\end{equation*}
We use this importance distribution in all of the case studies illustrated in Sect.~\ref{sec:applications}.

\subsubsection{Approximation of the Hessian with Louis' identity}
\label{sec:louis_identity}
Consider again the $i$th iteration. Louis' identity~\citep{cappe2005} for the Hessian $\nabla^2_{\btheta} \log p(\y_i \mid \btheta)$ is
\begin{equation}
\label{eq:louis_identity}
    - \nabla^2_{\btheta} \log p(\y_i \mid \btheta) = \nabla_{\btheta} \log p(\y_i \mid \btheta) \nabla_{\btheta} \log p(\y_i \mid \btheta)^\top - \frac{\nabla^2_{\btheta} p(\y_i \mid \btheta)}{p(\y_i \mid \btheta)},
\end{equation}
where 
\begin{align}
    \frac{\nabla^2_{\btheta} p(\y_i \mid \btheta)}{p(\y_i \mid \btheta)} = &\int p(\balpha_i \mid \y_i, \btheta) \nabla_{\btheta} \log p(\y_i, \balpha_i \mid \btheta) \nabla_{\btheta} \log p(\y_i, \balpha_i \mid \btheta)^\top  \d \balpha_i \nonumber \\
    &+ \int p(\balpha_i \mid \y_i, \btheta) \nabla^2_{\btheta} \log p(\y_i, \balpha_i \mid \btheta)  \d \balpha_i \label{eq:louis_second_term}.
\end{align}

The first term on the right-hand side of~\eqref{eq:louis_identity} is obtained using Fisher's identity, as discussed in Sect. \ref{sec:fishers_identity}. The second term consists of two integrals (see~\eqref{eq:louis_second_term}), which can also be approximated using samples. Specifically, 
\begin{align*}
    \frac{\nabla^2_{\btheta} p(\y_i \mid \btheta)}{p(\y_i \mid \btheta)} 
    \approx & \frac{1}{S_\alpha} \sum_{s = 1}^{S_\alpha} \left( \nabla_{\btheta} \log p(\y_i, \balpha_i^{(s)} \mid \btheta) \nabla_{\btheta} \log p(\y_i, \balpha_i^{(s)} \mid \btheta)^\top + \nabla^2_{\btheta} \log p(\y_i, \balpha_i^{(s)} \mid \btheta) \right), 
\end{align*}
where $\balpha_i^{(s)} \sim p(\balpha_i \mid \y_i, \btheta)$ for $s = 1, \dots, S_\alpha$. If obtaining samples from $p(\balpha_i \mid \y_i, \btheta)$ is not straightforward, importance sampling (as in Sect.~\ref{sec:fishers_identity}) can be used instead. Following~\cite{nemeth2016particle}, for computational efficiency, we use the same samples $\{\balpha_i^{(s)}: s = 1, \dots, S_\alpha\}$ that were used to approximate the score using Fisher's identity and their corresponding normalised weights $\mathcal{W}_i$ to obtain the estimates of the second term in Louis' identity. Then
\begin{align*}
    \frac{\nabla^2_{\btheta} p(\y_i \mid \btheta)}{p(\y_i \mid \btheta)} \approx \sum_{s=1}^{S_\alpha} \bar{w}^{(s)}_i \left( \nabla_{\btheta} \log p(\y_i, \balpha_i^{(s)} \mid \btheta) \nabla_{\btheta} \log p(\y_i, \balpha_i^{(s)} \mid \btheta)^\top + \nabla^2_{\btheta} \log p(\y_i, \balpha_i^{(s)} \mid \btheta) \right).
\end{align*}
 
\subsection{Damped R-VGAL} 
\label{sec:damping}
A possible problem with R-VGAL is its instability in the first few observations, making it sensitive to the ordering of the observations. In Sect. \ref{sec:sensitivityRVGAL} of the online supplement, we run the R-VGAL algorithm on a dataset in its original order, and also on a random reordering of the observations, and find that the R-VGAL parameter estimates from these two runs differ. Figures \ref{fig:trajectories_S1000_Sa1000} and \ref{fig:trajectories_seed2023_S1000_Sa1000} in Sect. \ref{sec:sensitivityRVGAL} show that the first few observations can heavily influence the trajectory of the variational mean. Here, we propose a damping approach to stabilise the R-VGAL algorithm during the initial few steps. 

\Copy{damped_RVGAL}{In damped R-VGAL, the updates of the mean and precision matrix for each observation are split into $K$ steps, where $K$ is selected on a case by case basis. In each step, we multiply the gradient and the Hessian of $\log p(\y_i \mid \btheta)$ by a factor $a= \frac{1}{K}$ (which acts as a ``step size"), and then update the variational parameters $K$ times during the $i$th iteration. Intuitively, in this way, one observation is split into $K$ ``parts" and incorporated into the updates one part at a time. Using a smaller step size helps stabilise the R-VGAL algorithm, particularly for the first few observations. 
Sect. \ref{sec:sensitivityRVGAL} of the online supplement shows that damping the first few iterations makes the R-VGAL algorithm more robust to different orderings of the data.

The damped R-VGAL approach we present here is inspired by the so-called \textit{damped Newton's method}. In the case where the model is linear and the likelihood is Gaussian, the original R-VGA algorithm, upon which R-VGAL is based, can be shown to be equivalent to an online version of Newton's method; see Appendix 8.2 of~\cite{lambert2022recursive} for a proof. 
Newton's method seeks the minimiser of a continuously differentiable function 
$f: \mathbb{R}^d \to \mathbb{R}, d \in \mathbb{N}$,
by beginning with some starting value $\u_0 \in \mathbb{R}^d$ and sequentially minimising the quadratic approximation of the function $f(\cdot)$ around the current value in order to find the next value:
\begin{equation*}
    \u_{k+1} = \argmin_{\u} f(\u_k) + \nabla_{\u} f(\u_k)^\top (\u-\u_k) + \frac{1}{2} (\u - \u_k)^\top \nabla^2_{\u} f(\u_k)(\u - \u_k), \quad k = 0,1,2,... 
\end{equation*}
Provided that $\nabla^2_{\u} f(\u_k)$ is positive definite, the minimiser of $f(\cdot)$ is unique and can be computed iteratively as 
\begin{equation}
    \u_{k+1} = \u_{k} - (\nabla^2_{\u} f(\u_k))^{-1} \nabla_{\u} f(\u_k), \quad k = 0,1,2,...
\end{equation}
These iterations stop when $\norm{\nabla f(\u_{k+1})} \leq \epsilon_0$, where $\epsilon_0$ is some small tolerance parameter. Often, in practice, Newton's method is modified to include a step size $0 < \rho \leq 1$ to improve convergence:
\begin{equation}
    \u_{k+1} = \u_{k} - \rho (\nabla^2_{\u} f(\u_k))^{-1} \nabla_{\u} f(\u_k), \quad k = 0,1,2,...,
\end{equation}
resulting in the damped Newton's method. This step size $\rho$ is similar to the multiplicative factor $a$ in our damped R-VGAL approach.} 

\Copy{naturalgrad}{We also note that, in the case where the model is linear or when the likelihood function comes from an exponential family and the model is linearised, the R-VGA algorithm of~\cite{lambert2022recursive} is equivalent to an online natural gradient algorithm with step size $\frac{1}{1+t}$, where $t$ denotes the iteration. A proof of this equivalence can be found in Appendix 8.3 of~\cite{lambert2022recursive}. Viewed from the perspective of natural gradient optimisation, the damping factor $a$ in damped R-VGAL can be interpreted as a reduction of the step size in natural gradient updates.} 

We summarise the damped R-VGAL algorithm in Algorithm \ref{algo:rvga_temper}.


\begin{algorithm} [t]
\caption{Damped R-VGAL} 
\label{algo:rvga_temper}
\begin{algorithmic}
\setstretch{1.5}
    \State Input: observations $\y_1, \dots, \y_N$, initial values $\bmu_0$ and $\bSigma_0$, number of observations to damp $n_{damp}$, number of damping steps $K$.
    \State Output: variational parameters $\bmu_i$ and $\bSigma_i$, for $i=1,...,N$.
    \State Set $q_0(\btheta) = \Gau(\bmu_0, \bSigma_0)$.
    \For {$i = 1, \dots, N$}
        \If {$i \leq n_{damp}$} 
        \State Set $a = 1/K$, $\bmu_{i, 0} = \bmu_{i-1}, \bSigma_{i, 0} = \bSigma_{i-1}$ 
            \For {$k = 1, \dots, K$}
            \State Set $q_{i, k-1}(\btheta) = \Gau(\bmu_{i, k-1}, \bSigma_{i, k-1})$.
            \State $\bmu_{i, k} = \bmu_{i, k-1} + a \bSigma_{i, k} \Exp_{q_{i, k-1}} (\wh{{\nabla_{\btheta} \log p(\y_i \mid \btheta)}})$ 
            \State $\bSigma_{i, k}^{-1} = \bSigma_{i, k-1}^{-1} - a \Exp_{q_{i, k-1}} (\wh{\nabla_{\btheta}^2 \log p(\y_i \mid \btheta)})$ 
            \EndFor
            \State Set $\bmu_{i} = \bmu_{i, K}, \bSigma_{i} = \bSigma_{i, K}$, $q_{i}(\btheta) = \Gau(\bmu_i, \bSigma_i)$.
        \Else 
            \State $\bmu_i = \bmu_{i-1} + \bSigma_i \Exp_{q_{i-1}} (\wh{\nabla_{\btheta} \log p(\y_i \mid \btheta)})$ 
            \State $\bSigma_i^{-1} = \bSigma_{i-1}^{-1} - \Exp_{q_{i-1}} (\wh{\nabla_{\btheta}^2 \log p(\y_i \mid \btheta)})$ 
        \EndIf
    \EndFor
\end{algorithmic}
\end{algorithm}
\section{Applications of R-VGAL}
\label{sec:applications}

In this section, we apply R-VGAL to estimate parameters in linear, logistic and Poisson mixed models using three simulated datasets and two real datasets: the Six City dataset from~\cite{fitzmaurice1993likelihood}, and the Polypharmacy dataset from~\cite{hosmer2013applied}. The linear and logistic models have univariate random effects, while the Poisson model has bivariate random effects. There are two additional examples in Sect.~\ref{sec:add_examples} of the online supplement: a real data example with the Poisson model applied to the Epilepsy dataset from~\cite{thallvail1990some}, and a synthetic data example with a high number of observations simulated from the logistic mixed model. 

We validate R-VGAL against Hamiltonian Monte Carlo \citep[HMC,][]{neal2011, betancourt2015hamiltonian}, which is implemented using the Stan programming language~\citep{stan} in R~\citep{R}. In examples with real data, the true parameters are unknown. We instead compute the maximum likelihood estimates for the parameters using the R package \texttt{lme4}~\citep{lme4}, and also treat results from HMC as the ``ground truth", as HMC provides samples from the true posterior distributions. For all examples, we run 2 HMC chains for $15000$ iterations each, and discard the first $5000$ from each chain as burn in. We find that the effective sample sizes are high and the $\hat{R}$ statistics are close to 1 for all examples, indicating that the HMC chains are well-mixed and have converged; see Sect.~\ref{sec:hmc_convergence} of the online supplement for further details. Reproducible R code for all examples is available on \url{https://github.com/bao-anh-vu/R-VGAL}. 

For all applications in this paper, we use the damped R-VGAL algorithm described in Sect.~\ref{sec:damping}. We show that damping makes the algorithm more robust to different orderings of the observations in Sect.~\ref{sec:sensitivityRVGAL} of the online supplement. \Copy{n_temp}{The values of $n_{damp}$ and $K$ used in damping observations should be kept as small as possible to limit the extra computational overhead, while also be sufficiently large to reduce the instability observed with the R-VGAL algorithm in the initial stages. In our applications, we experimented with a few different settings of $n_{damp}$ and $K$ and plotted the trajectories of the variational mean under those settings. We found that the trajectories were most unstable during the first 10 observations, so we chose $n_{damp} = 10$ observations and the number of steps $K = 4$ to reduce the initial instability at the expense of a small additional computational cost. These values are used throughout our examples. Adaptive schemes for selecting the values of $n_{damp}$ and $K$ are left as future research directions.}


\subsection{Linear mixed effect model}
\label{sec:lmm}
In this example, we generate data from a linear mixed model with $N = 200$ groups and $n = 10$ responses per group. The $j$th response from the $i$th group is modelled as  
\begin{equation}
    y_{ij} = \x_{ij}^\top \bbeta + z_{ij} \alpha_i + \epsilon_{ij}, \quad \alpha_i \sim \Gau(0, \sigma_\alpha^2), \quad \epsilon_{ij} \sim \Gau(0, \sigma_\epsilon^2), 
\end{equation}
for $i = 1, \dots, N$ and $ j = 1, \dots, n$, where $\x_{ij}$ is drawn from a $\Gau(\0, \I_4)$ distribution and $z_{ij}$ is drawn from a $\Gau(0, 1)$ distribution. 
For this example, we did not include an intercept term, but it can be added if necessary. The true parameter values are $\bbeta = (-1.5, 1.5, 0.5, 0.25)^\top$, $\sigma_\alpha = 0.9$, and $\sigma_\epsilon = 0.7$. Since R-VGAL uses a multivariate normal distribution as the variational approximation, we consider the log-transformed variables $\phi_\alpha \equiv \log(\sigma_{\alpha}^2)$ and $\phi_\epsilon \equiv \log(\sigma_{\epsilon}^2)$ so that $\phi_\alpha$ and $\phi_\epsilon$ are unconstrained. We then make inference on the parameters $\btheta = (\bbeta^\top, \phi_\alpha, \phi_\epsilon)^\top$ using R-VGAL.

At the group level, the linear mixed model is 
\begin{equation}
    \y_i = \X_i \bbeta + \z_i \alpha_i + \bepsilon_i, \quad i = 1, \dots, N,
\end{equation}
where $\y_i \equiv (y_{i1}, \dots, y_{in})^\top$, $\X_i \equiv (\x_{i1}, \dots, \x_{in})^\top, \z_i \equiv (z_{i1}, \dots, z_{in})^\top$, and $\bepsilon_i \equiv (\epsilon_{i1}, \dots, \epsilon_{in})^\top$. At each iteration, $i = 1, \dots, N$, the R-VGAL algorithm makes use of the ``partial" likelihood of the observations from the $i$th group, $p(\y_i \mid \btheta) = \Gau(\bmu_{y \mid \theta}, \bSigma_{y \mid \theta})$, where $\bmu_{y \mid \theta} = \X_i \bbeta$ and $\bSigma_{y \mid \theta} = \sigma_\alpha^2 \z_i \z_i^\top + \sigma_{\epsilon}^2 \I_n$. For this model, the gradient and Hessian of $\log p(\y_i \mid \btheta)$ with respect to each of the parameters are available in closed form; see Sect. \ref{derivativelinearmodel} of the online supplement. In this case, we are therefore able to compare the accuracy of R-VGAL implemented using approximate gradients and Hessians with that of R-VGAL implemented using exact gradients and Hessians. 

The prior distribution we use, which is also the ``initial" variational distribution, is 
\begin{equation}
    p(\btheta) = q_0(\btheta) = \Gau \left(
    \begin{bmatrix} \0 \\ \log(0.5^2) \\ \log(0.5^2) \end{bmatrix}, 
    \begin{bmatrix} 10\I_4 & \0 & \0 \\
    \0^\top & 1 & 0 \\
    \0^\top & 0 & 1 \\
    \end{bmatrix}
    \right).
\end{equation}
A $\Gau(\log(0.5^2), 1)$ prior distribution for $\phi_\alpha$ and $\phi_\epsilon$ is equivalent to a log-normal prior distribution with mean 0.41 and variance 0.29 for both $\sigma_\alpha^2$ and $\sigma_\epsilon^2$. Using this prior distribution, the $2.5$th and $97.5$th percentiles for both $\sigma_\alpha^2$ and $\sigma_\epsilon^2$ are $(0.035, 1.775)$. 

At each iteration $i = 1, \dots, 200$, we use $S_\alpha = 100$ Monte Carlo samples (of $\alpha_i$) to approximate the gradient and Hessian of $\log p(\y_i \mid \btheta)$ using Fisher's and Louis' identities. We use $S = 100$ Monte Carlo samples (of $\btheta$) to approximate the expectations with respect to $q_{i-1}(\btheta)$ in the R-VGAL updates of the mean and precision matrix. These values were chosen based on an experimental study on the effect of $S$ and $S_\alpha$ on the posterior estimates of R-VGAL in Sect.~\ref{sec:var_test_S_Salpha} of the online supplement.

\begin{figure}[t]
    \centering
    \includegraphics[width = \linewidth]{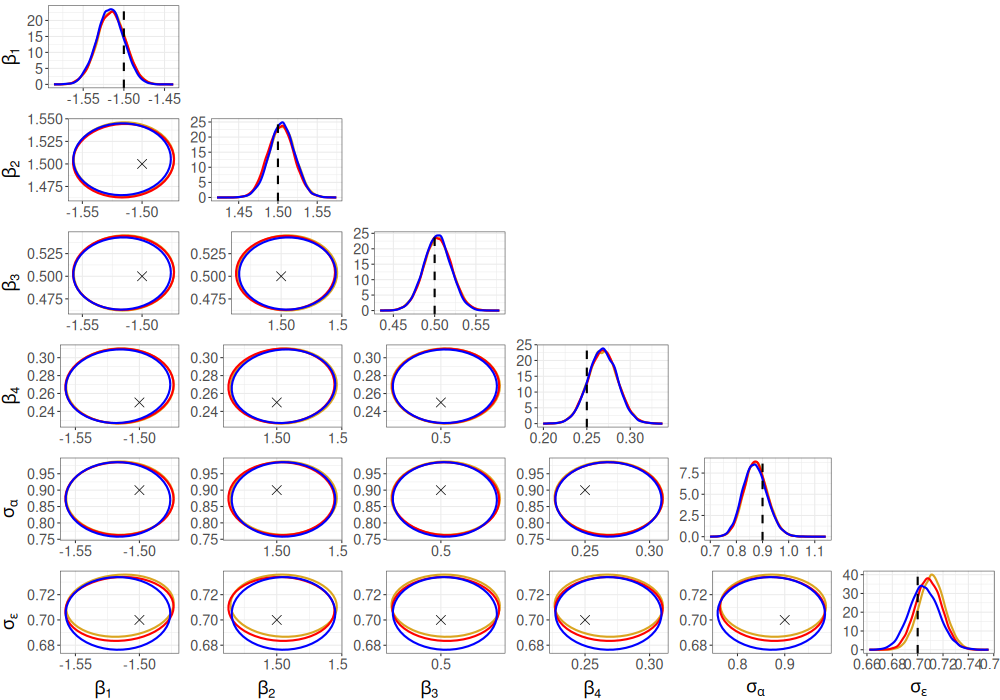}
    \caption{Exact posterior distributions (from HMC, in blue) and approximate posterior distributions (from R-VGAL with estimated gradients and Hessians in red, and from R-VGAL with exact gradients and Hessians in yellow) for the linear mixed model experiment. Diagonal panels: Marginal posterior distributions with true parameters denoted using dotted lines. Off-diagonal panels: Bivariate posterior distributions with true parameters denoted using the symbol $\times$.}
    \label{fig:linear_posterior}
\end{figure}

We validate R-VGAL against HMC, which we implemented in Stan. Figure~\ref{fig:linear_posterior} shows the marginal posterior distributions of the parameters, along with bivariate posterior distributions as estimated using R-VGAL with approximate gradients and Hessians, R-VGAL with exact gradients and Hessians, and HMC. The posterior distributions obtained using R-VGAL are clearly very similar to those obtained using HMC, irrespective of whether exact or approximate gradients and Hessians are used.

\subsection{Logistic mixed effect model}
\label{sec:logmm}
In this example, we generate simulated data from a random effects logistic regression model with $N = 500$ groups and $n = 10$ responses per group. The random effect logistic regression model we use is 
\begin{equation}
    y_{ij} \sim \textrm{Bernoulli} (\pi_{ij}), \quad \pi_{ij} \equiv \textrm{Pr}(y_{ij} = 1 \mid \bbeta, \tau^2) = \frac{\exp(\x_{ij}^\top \bbeta + \alpha_i)}{1 + \exp(\x_{ij}^\top \bbeta + \alpha_i)}, \quad \alpha_i \sim \Gau(0, \tau^2),
\end{equation}
where $\x_{ij}$ is drawn from a $\Gau(\0, \I_4)$ distribution, for $i = 1, \dots, N$  and $j = 1, \dots, n$. For this example, we did not include an intercept term, but it can be added if necessary. The true parameter values are $\bbeta = (-1.5, 1.5, 0.5, 0.25)^\top$ and $\tau = 0.9$.  

As in the linear case, although the parameters of the model are $\bbeta$ and $\tau$, we work with $\btheta = (\bbeta^\top, \phi_\tau)^\top$ where $\phi_\tau \equiv \log (\tau^2)$. The gradient and Hessian of the ``partial" log-likelihood $\log p(\y_i \mid \btheta)$ in this model are not analytically tractable, but can be estimated unbiasedly using Fisher's and Louis' identities as discussed in Sects.~\ref{sec:fishers_identity} and~\ref{sec:louis_identity}. These identities require the expressions for ${\nabla_{\btheta} \log p(\y_i, \alpha_i \mid \btheta)}$ and ${\nabla^2_{\btheta} \log p(\y_i, \alpha_i \mid \btheta)}$, which are provided in Sect. \ref{derivativelogisticregression} of the online supplement.

The prior distribution we use, which is also the ``initial" variational distribution, is
\begin{equation}
\label{initialvariationaldistlogistic}
    p(\btheta) = q_0(\btheta) = \Gau \left(
    \begin{bmatrix} \0 \\ \log(0.5^2) \end{bmatrix}, 
    \begin{bmatrix} 10\I_4 & \0\\
    \0^\top & 1
    \end{bmatrix}
    \right).
\end{equation}
A $\Gau(\log(0.5^2), 1)$ prior distribution for $\phi_\tau$ is equivalent to a log-normal prior distribution with mean 0.41 and variance 0.29 for $\tau^2$. The prior $2.5$th and $97.5$th percentiles for $\tau^2$ are $(0.035, 1.775)$.
At each iteration $i = 1, \dots, 500$, we use $S_\alpha = 100$ Monte Carlo samples (of $\alpha_i$) to approximate the gradient and Hessian of $\log p(\y_i \mid \btheta)$ during importance sampling, and $S = 100$ samples (of $\btheta$) to approximate the expectations with respect to $q_{i-1}(\btheta)$ in the R-VGAL updates of the mean and precision matrix.

\begin{figure}[t]
    \centering
    \includegraphics[width = 0.9\linewidth]{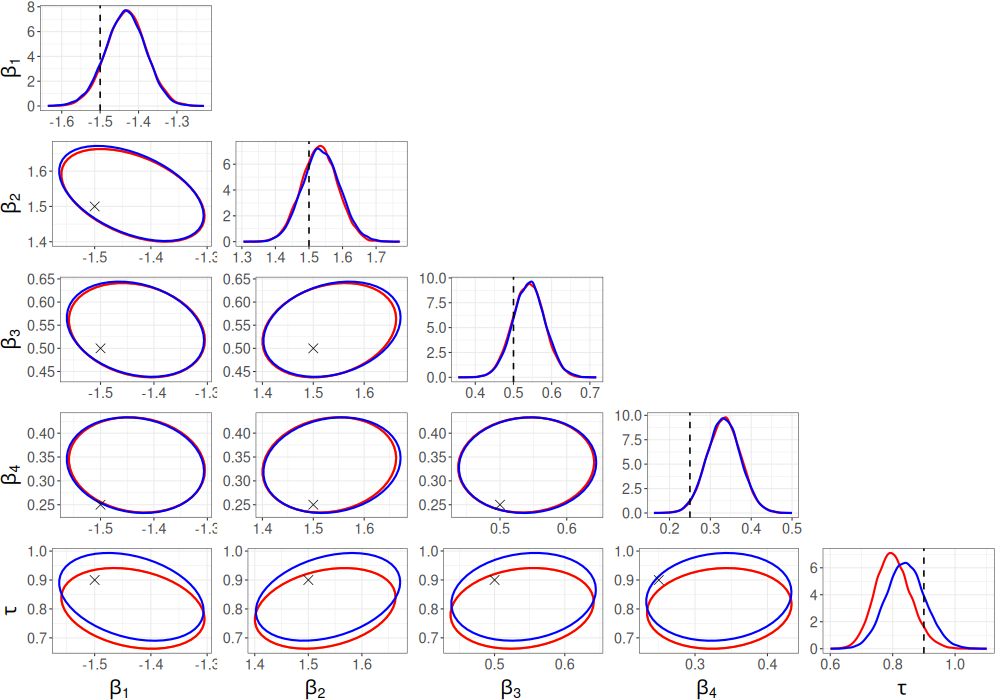}
    \caption{Exact posterior distributions from HMC (in blue) and approximate posterior distributions from R-VGAL with estimated gradients and Hessians (in red) for the logistic mixed model experiment. Diagonal panels: Marginal posterior distributions with true parameters denoted using dotted lines. Off-diagonal panels: Bivariate posterior distributions with true parameters denoted using the symbol $\times$.}
    \label{fig:logistic_posterior}
\end{figure}

Figure~\ref{fig:logistic_posterior} shows the marginal posterior distributions of the parameters, along with bivariate posterior distributions as estimated using R-VGAL and HMC. The posterior distributions obtained using R-VGAL are again very similar to those obtained using HMC.

\subsection{Poisson mixed model}
\label{sec:poisson}
We now apply R-VGAL to a model with bivariate random effects. For this example, we simulate data with $N = 200$ groups and $n = 10$ responses per group from the following Poisson mixed effect regression model:
\begin{equation*}
    y_{ij} \sim \text{Poisson} (\lambda_{ij}), \quad \lambda_{ij} = \exp(\x_{ij}^\top \bbeta + \z_{ij}^\top \balpha_i), \quad \balpha_i \sim \Gau(\0, \bSigma_\alpha),
\end{equation*}
where $\x_{ij} \equiv (1, x_{ij,1})^\top$, with $x_{ij, 1}$ drawn from a $\Gau(0, 1)$ distribution, and $\z_{ij} \equiv (1, z_{ij,1})^\top$, with $z_{ij,1}$ drawn from a $\Gau(0, 1)$ distribution, for $i = 1, \dots, N$ and $j = 1, \dots, n$. We denote the fixed and random effects as $\bbeta \equiv (\beta_0, \beta_1)^\top$ and $\balpha_i \equiv (\alpha_{i, 1}, \alpha_{i, 2})^\top$, respectively. The true parameter values are 
\begin{equation*}
    \bbeta = (-1.5, -0.5)^\top, \quad 
    \bSigma_\alpha = \begin{bmatrix} 
        0.15 & 0.05 \\
        0.05 & 0.20
    \end{bmatrix}.
\end{equation*}

We parameterise $\bSigma_\alpha = \L\L^\top$, where $\L$ denotes the lower Cholesky factor of $\bSigma_\alpha$ and takes the form
\begin{equation*}
    \L = \begin{bmatrix}
        \exp(\zeta_{11}) & 0 \\
        \zeta_{21} & \exp(\zeta_{22})  
    \end{bmatrix}.
\end{equation*}
In the algorithm, we consider the unconstrained parameters $\btheta = (\bbeta^\top, \bzeta^\top)^\top$,  where $\bzeta \equiv (\zeta_{11}, \zeta_{22}, \zeta_{21})^\top$. The gradient $\nabla_{\btheta} \log p(\y_i, \balpha_i \mid \btheta)$ and Hessian $\nabla^2_{\btheta} \log p(\y_i, \balpha_i \mid \btheta)$, which are necessary in the computation of the gradient and Hessian of the group-specific log likelihood $\log p(\y_i \mid \btheta)$, are provided in Sect.~\ref{derivativepoissonregression} of the online supplement. 

We use the following prior/initial variational distribution:
\begin{equation*}
    p(\btheta) = q_0(\btheta) = 
    \Gau \left(
    \begin{bmatrix} \0 \\ \0 \end{bmatrix}, 
    \begin{bmatrix} \I_2 & \0\\
    \0^\top & 0.1 \I_3
    \end{bmatrix}
    \right).
\end{equation*}
Using a $\Gau(0, 0.1)$ prior distribution for $\zeta_{11}$,  $\zeta_{22}$ and $\zeta_{21}$ leads to having 2.5th and 97.5th percentiles of ${(0.290, 3.485)}$ for $\Sigma_{\alpha_{11}}$, $(0.342, 3.577)$ for $\Sigma_{\alpha_{22}}$, and $(-0.713, 0.713)$ for the off-diagonal entries $\Sigma_{\alpha_{21}}$ and $\Sigma_{\alpha_{12}}$. 

\begin{figure}[t]
    \centering
    \includegraphics[width = 0.9\linewidth]{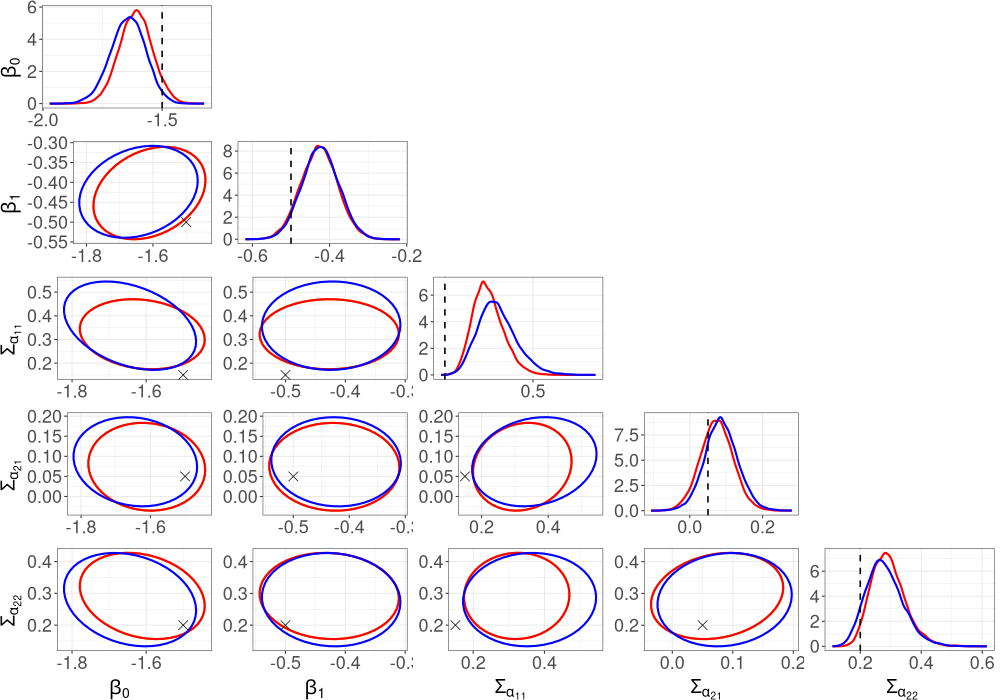}
    \caption{Exact posterior distributions from HMC (in blue) and approximate posterior distributions from R-VGAL with estimated gradients and Hessians (in red) for the Poisson mixed model experiment. Diagonal panels: Marginal posterior distributions with true parameters denoted using dotted lines. Off-diagonal panels: Bivariate posterior distributions with true parameters denoted using the symbol $\times$.}
    \label{fig:poisson_posterior_temper10_S100_Sa100_20231224}
\end{figure}

As with the linear and logistic examples, we use $S_\alpha = 100$ for the importance sampling step and $S = 100$ samples for approximating the expectations with respect to $q_{i-1}(\btheta)$ in the R-VGAL updates.
Figure~\ref{fig:poisson_posterior_temper10_S100_Sa100_20231224} shows the marginal posterior distributions of the parameters, along with bivariate posterior distributions as estimated using R-VGAL and HMC. For all parameters, the R-VGAL and HMC posterior densities are very similar, though the posterior densities of $\Sigma_{\alpha_{11}}$ from both methods appear a bit biased.

To assess the robustness of the results in these simulation studies, we also include repeated simulation studies on the linear, logistic and Poisson mixed models in Sect.~\ref{sec:multi_sims} of the online supplement. For each of these models, we simulate 100 datasets using the same parameter settings, and compare the posterior estimates from R-VGAL and HMC on these simulated datasets. We find that the R-VGAL and HMC posterior estimates are very similar across simulations for the linear and logistic models, while for the Poisson model, the estimates from the two methods are close for most simulations, with only a few cases where estimates are slightly different. We also find that the posterior standard deviations from R-VGAL tend to be slightly smaller than those from HMC.

\subsection{Real data examples}
\label{sec:realdata}
We now apply R-VGAL to two real datasets: the Six City dataset from~\cite{fitzmaurice1993likelihood}, and the Polypharmacy dataset from~\cite{hosmer2013applied}.

For the Six City dataset, we follow~\cite{tran2017variational} and consider the random intercept logistic regression model
\begin{align}
    \log \left(\frac{\pi_{ij}}{1 - \pi_{ij}} \right) = \beta_0 + \beta_{age} \texttt{Age}_{ij} + \beta_{smoke} \texttt{Smoke}_{ij} + \alpha_i, \quad \alpha_i \sim \Gau(0, \tau^2),
\end{align}
where $\pi_{ij} \equiv \textrm{Pr}(y_{ij} = 1 \mid \boldsymbol \beta, \tau^2)$, with $\bbeta \equiv (\beta_0, \beta_{age}, \beta_{smoke})^\top$, for $i = 1, \dots, 537$ and $j = 1, \dots, 4$. The binary response variable $y_{ij} = 1$ if child $i$ is wheezing at time point $j$, and 0 otherwise. The covariate $\texttt{Age}_{ij}$ is the age of child $i$ at time point $j$, centred at 9 years, while the covariate $\texttt{Smoke}_{ij} = 1$ if the mother of child $i$ is smoking at time point $j$, and 0 otherwise. Finally, $\alpha_i$ is the random effect associated with the $i$th child. The parameters of the model are $\btheta =
(\bbeta^\top, \phi_\tau)^\top$, where $\phi_\tau \equiv \log(\tau^2)$. 

For the Polypharmacy dataset, we consider the random intercept logistic regression model from~\cite{tan2018gaussian}:
\begin{align}
    \log \left(\frac{\pi_{ij}}{1 - \pi_{ij}} \right) = \beta_0 &+ \beta_{gender} \texttt{Gender}_{i} + \beta_{race} \texttt{Race}_{i} + \beta_{age} \texttt{Age}_{ij} + \beta_{M1} \texttt{MHV1}_{ij} \nonumber \\ 
    & + \beta_{M2} \texttt{MHV2}_{ij} + \beta_{M3} \texttt{MHV3}_{ij} + \beta_{IM} \texttt{INPTMHV}_{ij} + \alpha_i, \quad \alpha_i \sim \Gau(0, \tau^2),
\end{align}
where $\pi_{ij} \equiv \textrm{Pr}(y_{ij} = 1 \mid \boldsymbol\beta,\tau^2)$, $\bbeta \equiv (\beta_0, \beta_{gender}, \beta_{race}, \beta_{age}, \beta_{M1}, \beta_{M2}, \beta_{M3}, \beta_{IM})^\top$, for $i = 1, \dots, 500$ and $j = 1, \dots, 7$. The response variable $y_{ij}$ is 1 if subject $i$ in year $j$ is taking drugs from three or more different classes (of drugs), and 0 otherwise. The covariate $\texttt{Gender}_i = 1$ if subject $i$ is male, and $0$ if female, while $\texttt{Race}_i = 0$ if the race of subject $i$ is white, and $1$ otherwise. The covariate $\texttt{Age}_{ij}$ is the age (in years and months, to two decimal places) of subject $i$ in year $j$. The number of outpatient mental health visits (MHV) for subject $i$ in year $j$ is split into three dummy variables: $\texttt{MHV1}_{ij} = 1$ if $1 \leq \texttt{MHV}_{ij} \leq 5$, and $0$ otherwise; $\texttt{MHV2}_{ij} = 1$ if $6 \leq \texttt{MHV}_{ij} \leq 14$, and $0$ otherwise; and $\texttt{MHV3}_{ij} = 1$ if $\texttt{MHV}_{ij} \geq 15$, and $0$ otherwise. The covariate $\texttt{INPTMHV}_{ij} = 0$ if there were no inpatient mental health visits for subject $i$ in year $j$, and $1$ otherwise. Finally, $\alpha_i$ is a subject-level random effect for subject $i$. The parameters of the model are $\btheta =
(\bbeta^\top, \phi_\tau)^\top$, where $\phi_\tau \equiv \log(\tau^2)$. 

We use similar priors/initial variational distributions for both examples. For the Six City dataset, the prior/initial variational distribution we use is 
\begin{equation}
    p(\btheta) = q_0(\btheta) = \Gau \left(
    \begin{bmatrix} \0 \\ 1 \end{bmatrix}, 
    \begin{bmatrix} 10\I_3 & \0 \\
    \0^\top & 1 \end{bmatrix}
    \right),
\end{equation}
and for the Polypharmacy dataset, we use
\begin{equation}
    p(\btheta) = q_0(\btheta) = \Gau \left(
    \begin{bmatrix} \0 \\ 1 \end{bmatrix}, 
    \begin{bmatrix} 10\I_8 & \0 \\
    \0^\top & 1 \end{bmatrix}
    \right).
\end{equation}
A $\Gau(1,1)$ prior distribution for $\phi_\tau$ leads to a log-normal prior distribution with mean 4.48 and variance 34.51 for $\tau^2$. Using this prior distribution, the $2.5$th and $97.5$th percentiles for $\tau^2$ are $(0.383, 19.297)$, which cover most values of $\tau^2$ in practice. 
At each R-VGAL iteration, the gradient and Hessian of ${\log p(\y_i \mid \btheta)}$ are approximated using $S_\alpha = 200$ Monte Carlo samples (of $\alpha_i$), and the expectations with respect to $q_{i-1}(\btheta)$ in the R-VGAL updates are approximated using $S = 200$ Monte Carlo samples (of $\btheta$). 

\begin{figure}[t]
    \centering
    \includegraphics[width = 0.8\linewidth]{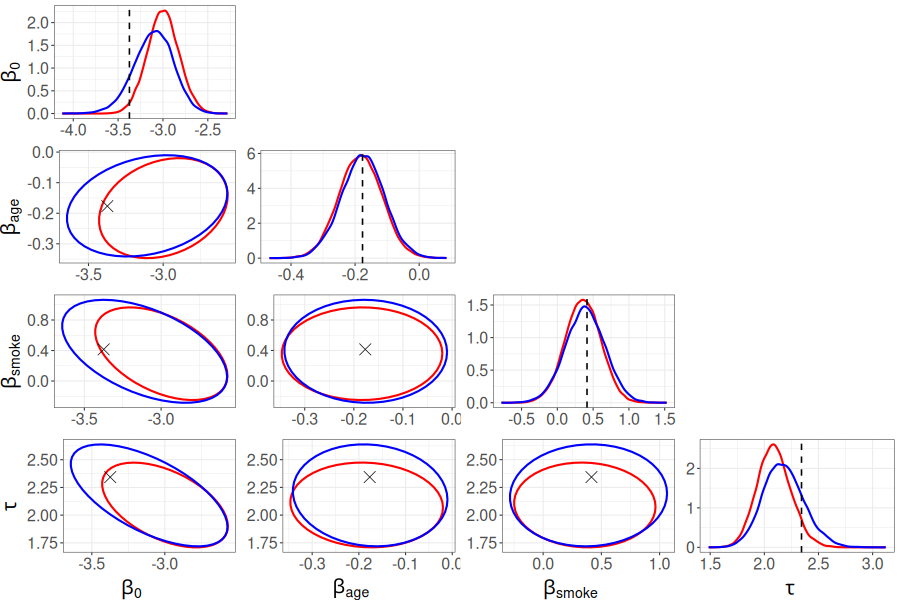}
    \caption{Exact posterior distributions from HMC (in blue) and approximate posterior distributions from R-VGAL with estimated gradients and Hessians (in red) for the experiment with the Six City dataset. Diagonal panels: Marginal posterior distributions with the maximum likelihood estimates marked using dotted lines. Off-diagonal panels: Bivariate posterior distributions with the maximum likelihood estimates marked using the symbol $\times$.}
    \label{fig:sixcity_posterior}
\end{figure}

\begin{figure}[t]
    \centering
    \includegraphics[width = \linewidth]{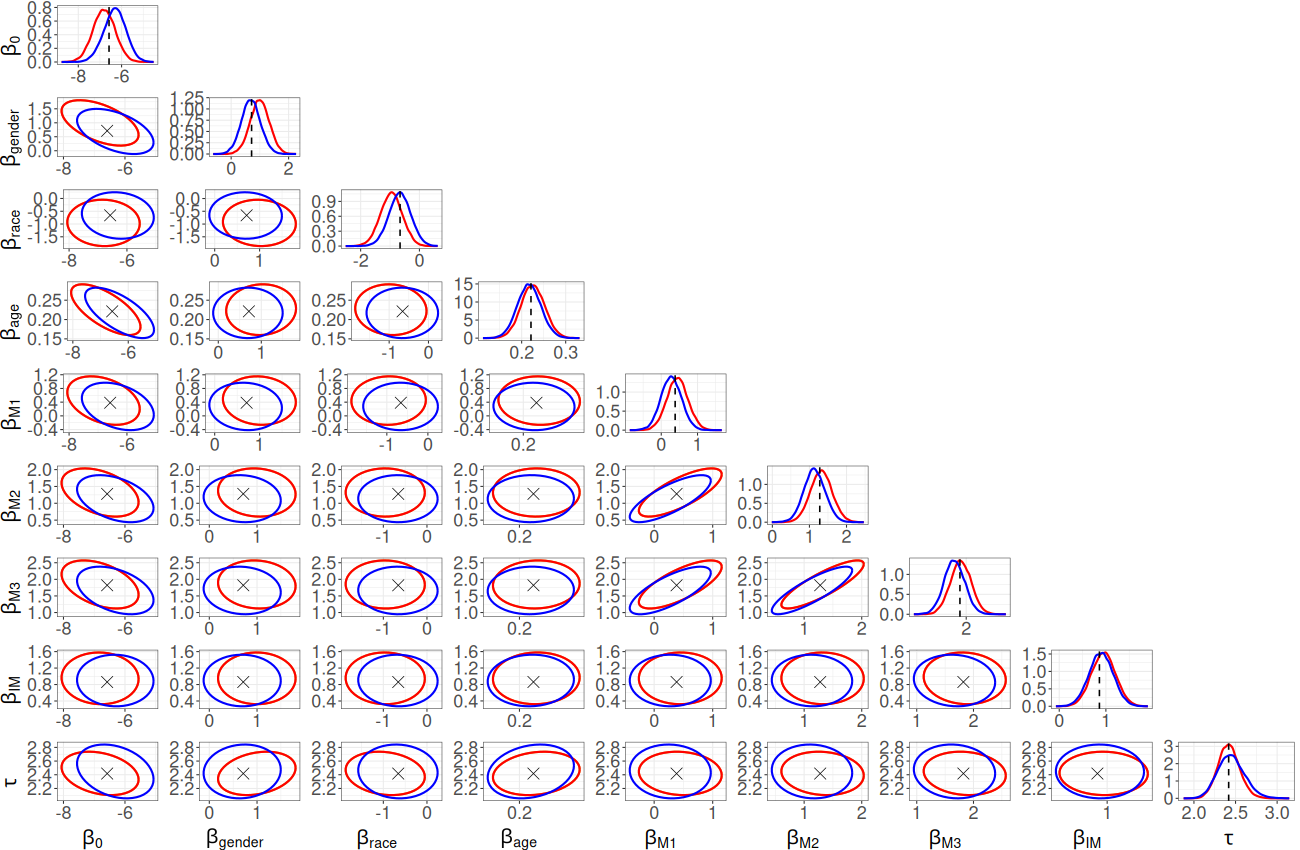}
    \caption{Exact posterior distributions from HMC (in blue) and approximate posterior distributions from R-VGAL with estimated gradients and Hessians (in red) for the experiment with the Polypharmacy dataset. Diagonal panels: Marginal posterior distributions  with the maximum likelihood estimates marked using dotted lines. Off-diagonal panels: Bivariate posterior distributions  with the maximum likelihood estimates marked using the symbol $\times$.}
    \label{fig:polypharmacy_posterior}
\end{figure}

As there are no ground truths to these examples, we compare the posterior density estimates from R-VGAL to those from HMC. In addition, we also compute the maximum likelihood estimates using the \texttt{lme4} package in R. Figures~\ref{fig:sixcity_posterior} and~\ref{fig:polypharmacy_posterior} show the marginal posterior distributions with maximum likelihood estimates of the parameters, along with bivariate posterior distributions estimated using R-VGAL and HMC for the Six City and Polypharmacy datasets, respectively. In the Six City example, there is a slight difference in the marginal and bivariate posterior densities from R-VGAL and HMC for the fixed effect $\beta_{smoke}$, but the posterior densities for other parameters are very similar between the two methods. For the intercept $\beta_0$ and the random effect standard deviation $\tau$, the posterior modes of HMC are closer to the maximum likelihood estimates than the posterior modes of R-VGAL, but for the other parameters, the posterior modes from both R-VGAL and HMC are close to the maximum likelihood estimates. For the Polypharmacy example, there are slight differences between the R-VGAL and HMC marginal and bivariate posterior densities for the intercept $\beta_0$ and the fixed effects $\beta_{gender}$ and $\beta_{race}$, but for other parameters, the posterior densities are comparable between the two methods. The posterior modes of both R-VGAL and HMC are close to the maximum likelihood estimates for all parameters in this example.

\subsection{Computing time \label{sec:computing_time}}
Table~\ref{table:elapsed_time} compares the computing time (in minutes) of R-VGAL and HMC for all simulated and real data examples that we have discussed in Sect.~\ref{sec:applications} and Sect.~\ref{sec:add_examples} of the online supplement, and includes the corresponding dataset size for each example. The last column in the table shows the average time taken (in seconds) for a single iteration of R-VGAL. For the linear example, where we run R-VGAL with both the theoretical and estimated gradients/Hessians, the displayed time is that of R-VGAL with the estimated gradients/Hessians. 
\Copy{GPU}{All experiments were carried out on the High Performance Computer system of the National Institute for Applied Statistics Research Australia, with an NVIDIA Tesla V100 PCIe 32GB graphics processing unit (GPU). 
The GPU was used to parallelise the computations in the importance sampling step, so that the gradient and Hessian of the joint log-likelihood $\log p(\y_i, \balpha_i^{(s)} \mid \btheta), s = 1, \dots, S_\alpha$, and their corresponding weights $\mathcal{W}_i$, are computed all at once. The GPU was also used to parallelise over the Monte Carlo samples used in the estimation of the expectations with respect to $q_{i-1}(\cdot)$ in Algorithm~\ref{algo:rvgal}. We use the R interface to Tensorflow~\citep{tensorflow2015-whitepaper} to facilitate GPU computations.}

The table shows that \Copy{computing_time}{the R-VGAL algorithm is generally 3 to 8 times faster than HMC. 
This is substantial given that our code is not as highly optimised as that in Stan. The difference in computing times also becomes more notable with a bigger dataset: in the logistic example with $50000$ synthetic observations (see Sect.~\ref{sec:big_data} of the online supplement), R-VGAL takes only 17 minutes to produce posterior estimates, while HMC takes more than 2 hours.}
Furthermore, since R-VGAL is a sequential algorithm, posterior approximations from R-VGAL can be easily updated as new observations become available. To incorporate an additional observation, R-VGAL needs to  perform a single update, while an algorithm like HMC requires rerunning the entire sampling procedure. 



\begin{table}[t]
\centering
\begin{tabular}{|l|r|r|r|r|r|r}
\cline{1-6}
                  & $N$ & $n$ & HMC (min) & R-VGAL (min) & One R-VGAL iteration (s) &  \\ \cline{1-6}
Linear (simulated data)  & 200 & 10    &  2.5   &  0.6       &  0.17                      &  \\ \cline{1-6}
Logistic (simulated data) & 500 & 10   &  7.2   &  1.1    & 0.13                        &  \\ \cline{1-6}
Poisson (simulated data) & 200 & 10 & 11.3   &  3.1       &  1.05                     &  \\ \cline{1-6}
Logistic (Six City) & 537 & 4 & 3.4    &   1.2      &  0.13                     &  \\ \cline{1-6}
Logistic (Polypharmacy) & 500 & 7 &  18.5   &  2.4      &  0.29                     &  \\ \cline{1-6}
Poisson (Epilepsy)* & 59 & 4 & 3.3   &  1.2       &  1.25                     &  \\ \cline{1-6}
Logistic (simulated data)* & 5000 & 10 & 133.6   &  16.8 &  0.20                     &  \\ \cline{1-6}

\end{tabular}
\caption{Computing time (in minutes) for R-VGAL and HMC on simulated and real datasets, with accompanying dataset sizes. Timings for one R-VGAL update is shown (in seconds). Examples with the * symbol are in the online supplement.}
\label{table:elapsed_time}
\end{table}



\section{Conclusion}
\label{sec:conclusion}

In this article, we propose a sequential variational Bayes algorithm for estimating parameters in GLMMs based on an extension of the R-VGA algorithm of \citet{lambert2022recursive}. The original R-VGA algorithm requires the gradient and Hessian of the partial log-likelihood at each observation, which are computationally intractable for most GLMMs. To overcome this, we use Fisher's and Louis' identities to obtain unbiased estimates of the gradient and Hessian, which can be used in place of the closed form gradient and Hessian in the R-VGAL algorithm.

We apply R-VGAL to the linear, logistic and Poisson mixed effect models with simulated and real datasets. In all examples, we compare the posterior distributions of the parameters estimated using R-VGAL to those obtained using HMC~\citep{neal2011, betancourt2015hamiltonian}. The examples show that R-VGAL yields comparable posterior estimates to HMC while being substantially faster, and the R-VGAL posterior modes are very close to the maximum likelihood estimates for most parameters in the models we consider. R-VGAL would be especially useful in situations where new observations are being continuously collected.

In the current paper, we assume that the random effects are independent and identically distributed between subjects or groups. We discuss the potential application of R-VGAL to models with more complicated random effect structures, such as crossed or nested effects, in Sect.~\ref{sec:crossed_nested} of the online supplement. Future work will attempt to extend R-VGAL to cases where the random effects are temporally correlated. This will expand the set of models on which R-VGAL can be used to include time series and state space models.

\section*{Acknowledgements}

This work was supported by ARC SRIEAS Grant SR200100005 Securing Antarctica’s Environmental Future. Andrew Zammit-Mangion’s research was also supported by ARC Discovery Early Career Research Award DE180100203. The authors would also like to thank two anonymous reviewers for their insightful and constructive comments.

\bibliographystyle{apalike}
\bibliography{bibliography}

\newpage

    \renewcommand{\thealgorithm}{S\arabic{algorithm}}
    \renewcommand{\theequation}{S\arabic{equation}}
    \renewcommand{\thesection}{S\arabic{section}}
    \renewcommand{\thepage}{S\arabic{page}}
    \renewcommand{\thetable}{S\arabic{table}}
    \renewcommand{\thefigure}{S\arabic{figure}}
    \setcounter{page}{1}
    \setcounter{section}{0}
    \setcounter{equation}{0}
    \setcounter{algorithm}{0}
    \setcounter{table}{0}
    \setcounter{figure}{0}
    \clearpage 
\section{Appendix A: Gradient and Hessian derivations}

\subsection{Derivation of gradient and Hessian expressions for the linear mixed model\label{derivativelinearmodel}}

In the linear mixed model, the theoretical gradient and Hessian for the likelihood at observation $i$ for $i = 1, \dots, N$ are available. This section gives the expressions for both the theoretical gradient and Hessian, as well as their approximations when using Fisher's and Louis' identities.

\subsubsection{Theoretical gradient and Hessian}

The ``partial" log-likelihood for observations from the $i$th group is given by
\begin{equation}
    \log p(\y_i \mid \btheta) = -\frac{n}{2} \log (2\pi) - \frac{1}{2} \log \abs{{\bSigma_{y \mid \theta}}} - \frac{1}{2} (\y_i - \X_i \bbeta)^\top \bSigma_{y \mid \theta}^{-1}  (\y_i - \X_i \bbeta), \quad i = 1, \dots, N,
\end{equation}
where $\bSigma_{y \mid \theta} = \exp(\phi_\alpha) \z_i \z_i^\top + \exp(\phi_\epsilon) \I_n$.

The gradient of the log-likelihood with respect to the parameters $\btheta$ is given by 
\begin{equation*}
    \nabla_{\btheta} \log p(\y_i \mid \btheta) = (\nabla_{\bbeta} \log p(\y_i \mid \btheta)^\top, \nabla_{\phi_\alpha} \log p(\y_i \mid \btheta), \nabla_{\phi_\epsilon} \log p(\y_i \mid \btheta) )^\top,
\end{equation*}
where the components are, respectively,
\begin{align}
    \nabla_{\bbeta} \log p(\y_i \mid \btheta) &= \X_i^\top \bSigma_{y \mid \theta}^{-1} (\y_i - \X_i \bbeta) \label{eq:lmm_score_beta} \\
    \nabla_{\phi_\alpha} \log p(\y_i \mid \btheta) &= - \frac{1}{2} \Tr \left(\bSigma_{y \mid \theta}^{-1} \frac{\partial \bSigma_{y \mid \theta}}{\partial \phi_\alpha} \right) + \frac{1}{2} (\y_i - \X_i \bbeta)^\top \bSigma_{y \mid \theta}^{-1} \frac{\partial \bSigma_{y \mid \theta}}{\partial \phi_\alpha} \bSigma_{y \mid \theta}^{-1} (\y_i - \X_i \bbeta) \label{eq:lmm_score_phi} \\
    \nabla_{\phi_\epsilon} \log p(\y_i \mid \btheta) &= - \frac{1}{2} \Tr \left(\bSigma_{y \mid \theta}^{-1} \frac{\partial \bSigma_{y \mid \theta}}{\partial \phi_\epsilon} \right) + \frac{1}{2} (\y_i - \X_i \bbeta)^\top \bSigma_{y \mid \theta}^{-1} \frac{\partial \bSigma_{y \mid \theta}}{\partial \phi_\epsilon} \bSigma_{y \mid \theta}^{-1} (\y_i - \X_i \bbeta) \label{eq:lmm_score_psi},
\end{align}
with 
\begin{equation}
    \frac{\partial \bSigma_{y \mid \theta}}{\partial \phi_\alpha} = \exp(\phi_\alpha) \z_i \z_i^\top, \quad 
    \frac{\partial \bSigma_{y \mid \theta}}{\partial \phi_\epsilon} = \exp(\phi_\epsilon) \I_n.
\end{equation}
The Hessian is given by
\begin{equation}
    \nabla^2_{\btheta} \log p(\y_i \mid \btheta) = \begin{bmatrix}
    \nabla^2_{\bbeta} \log p(\y_i \mid \btheta) & \nabla_{\bbeta} \nabla_{\phi_\alpha} \log p(\y_i \mid \btheta) & \nabla_{\bbeta} \nabla_{\phi_\epsilon} \log p(\y_i \mid \btheta) \\
    \nabla_{\bbeta} \nabla_{\phi_\alpha} \log p(\y_i \mid \btheta)^\top & \nabla^2_{\phi_\alpha} \log p(\y_i \mid \btheta) & \nabla_{\phi_\alpha} \nabla_{\phi_\epsilon} \log p(\y_i \mid \btheta)\\
    \nabla_{\bbeta} \nabla_{\phi_\epsilon} \log p(\y_i \mid \btheta)^\top & \nabla_{\phi_\alpha} \nabla_{\phi_\epsilon} \log p(\y_i \mid \btheta) & \nabla^2_{\phi_\epsilon} \log p(\y_i \mid \btheta) \\
    \end{bmatrix}.
\end{equation}
The diagonal terms in the Hessian are the second derivatives
\begin{align}
    \nabla^2_{\bbeta} \log p(\y_i \mid \btheta) &= - \X_i^\top \bSigma_{y \mid \theta}^{-1} \X_i, \label{eq:grad2_beta} \\
    \nabla^2_{\phi_\alpha} \log p(\y_i \mid \btheta) &= -\frac{1}{2} \Tr(\G_{\phi_\alpha}) + \frac{1}{2} (\y_i - \X_i \bbeta)^\top \H_{\phi_\alpha} (\y_i - \X_i \bbeta), \label{eq:grad2_phi}   
\end{align}
where 
\begin{align}
\label{eq:lmm_hessian}
    \G_{\phi_\alpha} &= -\bSigma_{y \mid \theta}^{-1} \frac{\partial \bSigma_{y \mid \theta}}{\partial \phi_\alpha} \bSigma_{y \mid \theta}^{-1} \frac{\partial \bSigma_{y \mid \theta}}{\partial \phi_\alpha} + \bSigma_{y \mid \theta}^{-1} \frac{\partial^2 \bSigma_{y \mid \theta}}{\partial \phi_\alpha^2}\\
    \H_{\phi_\alpha} &= -2 \bSigma_{y \mid \theta}^{-1} \frac{\partial \bSigma_{y \mid \theta}}{\partial \phi_\alpha}\bSigma_{y \mid \theta}^{-1} \frac{\partial \bSigma_{y \mid \theta}}{\partial \phi_\alpha} \bSigma_{y \mid \theta}^{-1} + \bSigma_{y \mid \theta}^{-1} \frac{\partial^2 \bSigma_{y \mid \theta}}{\partial \phi_\alpha^2} \bSigma_{y \mid \theta}^{-1},
\end{align}
with $\frac{\partial^2 \bSigma_{y \mid \theta}}{\partial \phi_\alpha^2} = \exp(\phi_\alpha) \z_i \z_i^\top$. The expression for $\nabla^2_{\phi_\epsilon} \log p(\y_i \mid \btheta)$ is the same as in~\eqref{eq:grad2_phi}, but with all derivatives with respect to $\phi_\alpha$ replaced by those with respect to $\phi_\epsilon$. Note that $\frac{\partial^2 \bSigma_{y \mid \theta}}{\partial \phi_\epsilon^2} = \exp(\phi_\epsilon) \I_n$.

The off-diagonal terms in the Hessian are
\begin{align*}
    \nabla_{\bbeta} \nabla_{\phi_\alpha} \log p(\y_i \mid \btheta) &= -\X_i^\top \bSigma_{y \mid \theta}^{-1} \frac{\partial \bSigma_{y \mid \theta}}{\partial \phi_\alpha}  \bSigma_{y \mid \theta}^{-1} (\y_i - \X_i \bbeta) \\
    \nabla_{\bbeta} \nabla_{\phi_\epsilon} \log p(\y_i \mid \btheta) &= -\X_i^\top \bSigma_{y \mid \theta}^{-1} \frac{\partial \bSigma_{y \mid \theta}}{\partial \phi_\epsilon}  \bSigma_{y \mid \theta}^{-1} (\y_i - \X_i \bbeta)\\
    \nabla_{\phi_\alpha} \nabla_{\phi_\epsilon} \log p(\y_i \mid \btheta) &= -\frac{1}{2} \Tr(\G_{\phi_\alpha \phi_\epsilon}) + \frac{1}{2} (\y_i - \X_i \bbeta)^\top \H_{\phi_\alpha \phi_\epsilon} (\y_i - \X_i \bbeta), \label{eq:grad2_phi_psi}   
\end{align*}
where 
\begin{align}
    \G_{\phi_{\alpha} \phi_{\epsilon}} &= -\bSigma_{y \mid \theta}^{-1} \frac{\partial \bSigma_{y \mid \theta}}{\partial \phi_\alpha} \bSigma_{y \mid \theta}^{-1} \frac{\partial \bSigma_{y \mid \theta}}{\partial \phi_\epsilon} \\
    \H_{\phi_{\alpha} \phi_{\epsilon}} &= -\bSigma_{y \mid \theta}^{-1} \frac{\partial \bSigma_{y \mid \theta}}{\partial \phi_\alpha} \bSigma_{y \mid \theta}^{-1} \frac{\partial \bSigma_{y \mid \theta}}{\partial \phi_\epsilon} \bSigma_{y \mid \theta}^{-1} - \bSigma_{y \mid \theta}^{-1} \frac{\partial \bSigma_{y \mid \theta}}{\partial \phi_\epsilon} \bSigma_{y \mid \theta}^{-1} \frac{\partial \bSigma_{y \mid \theta}}{\partial \phi_\alpha} \bSigma_{y \mid \theta}^{-1}. 
\end{align}

\subsubsection{Expressions for the gradient and Hessian using Fisher's and Louis' identities}

Fisher's identity~\eqref{eq:fishers_identity} requires the gradient $\nabla_{\btheta} \log p(\y_i, \alpha_i \mid \btheta)$, while Louis' identity~\eqref{eq:louis_identity} requires the Hessian $\nabla_{\btheta}^2 \log p(\y_i, \alpha_i \mid \btheta)$. We now give the expression for these quantities for the linear mixed model considered in Sect.~\ref{sec:lmm}.

For $i = 1, \dots, N$, the gradient $\nabla_{\btheta} \log p(\y_i, \alpha_i \mid \btheta)$ can be written as
\begin{align}
    \nabla_{\btheta} \log p(\y_i, \alpha_i \mid \btheta) &= \nabla_{\btheta} \log p(\y_i \mid \alpha_i, \btheta) + \nabla_{\btheta} \log p(\alpha_i \mid \btheta), \\
    &= \nabla_{\btheta} \log p(\y_i \mid \alpha_i, \bbeta, \phi_\epsilon) + \nabla_{\btheta} \log p(\alpha_i \mid \phi_\alpha),
\end{align}
where 
\begin{equation}
    \log p(\y_i \mid \alpha_i, \bbeta, \phi_\epsilon) = -\frac{n}{2} \log(2\pi) - \frac{1}{2} \log \abs {\exp(\phi_\epsilon) \I_n} - \frac{1}{2 \exp(\phi_\epsilon)} (\y_i - \X_i \bbeta - \z_i \alpha_i)^\top (\y_i - \X_i \bbeta - \z_i \alpha_i),
\end{equation}
and 
\begin{equation}
    \log p(\alpha_i \mid \phi_\alpha) = - \frac{1}{2} \log (2\pi) - \frac{\phi_\alpha}{2} - \frac{\alpha_i^2}{2 \exp(\phi_\alpha)}.
\end{equation}
Thus, the gradient of the joint $\nabla_{\btheta} \log p(\y_i, \alpha_i \mid \btheta)$ is
\begin{equation}
    \nabla_{\btheta} \log p(\y_i, \alpha_i \mid \btheta) = \left(\nabla_{\bbeta} \log p(\y_i, \alpha_i \mid \btheta)^\top, \nabla_{\phi_\alpha} \log p(\y_i, \alpha_i \mid \btheta), \nabla_{\phi_\epsilon} \log p(\y_i, \alpha_i \mid \btheta) \right)^\top,
\end{equation}
where each component is given by
\begin{align}
    \nabla_{\bbeta} \log p(\y_i, \alpha_i \mid \btheta) 
    &= \nabla_{\bbeta} \log p(\y_i \mid \alpha_i, \bbeta, \phi_\epsilon) \nonumber \\ 
    &=\frac{1}{\exp(\phi_\epsilon)} \X_i^\top (\y_i - \X_i \bbeta - \z_i \alpha_i), \\
    \nabla_{\phi_\alpha} \log p(\y_i, \alpha_i \mid \btheta) &= \nabla_{\phi_\alpha} \log p(\alpha_i \mid \phi_\alpha) \nonumber \\
    &= - \frac{1}{2} + \frac{\alpha_i^2}{2 \exp(\phi_\alpha)}, \\
    \nabla_{\phi_\epsilon} \log p(\y_i, \alpha_i \mid \btheta) &= \nabla_{\phi_\epsilon} \log p(\y_i \mid \alpha_i, \bbeta, \phi_\epsilon) \nonumber \\
    &= - \frac{n}{2} + \frac{1}{2 \exp(\phi_\epsilon)} (\y_i - \X_i \bbeta - \z_i \alpha_i)^\top (\y_i - \X_i \bbeta - \z_i \alpha_i).
\end{align}
Similarly, the approximation of the Hessian using Louis' identity requires $\nabla^2_{\btheta} \log p(\y_i, \alpha_i \mid \btheta)$, in particular,
\begin{equation}
\setlength\arraycolsep{1pt}
\label{eq:lmm_joint_hessian}
    \nabla^2_{\btheta} \log p(\y_i, \alpha_i \mid \btheta) = \begin{bmatrix}
    \nabla^2_{\bbeta} \log p(\y_i, \alpha_i \mid \btheta) & \nabla_{\bbeta} \nabla_{\phi_\alpha} \log p(\y_i, \alpha_i \mid \btheta) & \nabla_{\bbeta} \nabla_{\phi_\epsilon} \log p(\y_i, \alpha_i \mid \btheta) \\
    \nabla_{\bbeta} \nabla_{\phi_\alpha} \log p(\y_i, \alpha_i \mid \btheta)^\top & \nabla^2_{\phi_\alpha} \log p(\y_i, \alpha_i \mid \btheta) & \nabla_{\phi_\alpha} \nabla_{\phi_\epsilon} \log p(\y_i, \alpha_i \mid \btheta)\\
    \nabla_{\bbeta} \nabla_{\phi_\epsilon} \log p(\y_i, \alpha_i \mid \btheta)^\top & \nabla_{\phi_\alpha} \nabla_{\phi_\epsilon} \log p(\y_i, \alpha_i \mid \btheta)^\top & \nabla^2_{\phi_\epsilon} \log p(\y_i, \alpha_i \mid \btheta) \\
    \end{bmatrix}.
\end{equation}
The components of~\eqref{eq:lmm_joint_hessian} are given by
\begin{align}
    \nabla^2_{\bbeta} \log p(\y_i, \alpha_i \mid \btheta) &= - \frac{1}{\exp(\phi_\epsilon)} \X_i^\top \X_i, \\
    \nabla^2_{\phi_\alpha} \log p(\y_i, \alpha_i \mid \btheta) &= - \frac{\alpha_i^2}{2 \exp(\phi_\alpha)}, \\
    \nabla^2_{\phi_\epsilon} \log p(\y_i, \alpha_i \mid \btheta) &= -\frac{1}{2 \exp(\phi_\epsilon)} (\y_i - \X_i \bbeta - \z_i \alpha_i)^\top (\y_i - \X_i \bbeta - \z_i \alpha_i),  \\
    \nabla_{\bbeta} \nabla_{\phi_\alpha} \log p(\y_i, \alpha_i \mid \btheta) &= \0, \\
    \nabla_{\bbeta} \nabla_{\phi_\epsilon} \log p(\y_i, \alpha_i \mid \btheta) &= -\frac{1}{\exp(\phi_\epsilon)} \X_i^\top (\y_i - \X_i \bbeta - \z_i \alpha_i), \\
    \nabla_{\phi_\alpha} \nabla_{\phi_\epsilon} \log p(\y_i, \alpha_i \mid \btheta) &= 0.
\end{align}

\subsection{Derivation of gradient and Hessian expressions for the logistic mixed model\label{derivativelogisticregression}}

The parameters of interest are $\btheta = (\bbeta^\top, \phi_\tau)^\top$, where $\phi_\tau = \log (\tau^2)$. The likelihood function $p(\y_i \mid \btheta)$ is not available in closed form for this model, so the gradient and Hessian of the log likelihood with respect to the parameters $\btheta$ need to be approximated via Fisher's and Louis' identities. 

The evaluation of Fisher's identity requires the gradient $\nabla_{\btheta} \log p(\y_i, \alpha_i \mid \btheta)$, where
\begin{align*}
    \log p(\y_i, \alpha_i \mid \btheta) &= \log p(\y_i \mid \alpha_i, \btheta) + \log p(\alpha_i \mid \btheta) \\
    &= \log p(\y_i \mid \alpha_i, \bbeta) + \log p(\alpha_i \mid \phi_\tau).
\end{align*}
Individually, 
\begin{equation*}
    \log p(\y_i \mid \alpha_i, \bbeta) = \sum_{j=1}^n y_{ij} \log \left( \frac{1}{1 + \exp(-(\x_{ij}^\top \bbeta + \alpha_i))} \right) + (1-y_{ij}) \log \left(1 - \frac{1}{1 + \exp(-(\x_{ij}^\top \bbeta + \alpha_i))} \right)
\end{equation*}
and 
\begin{equation*}
    \log p(\alpha_i \mid \phi_\tau) = -\frac{1}{2} \log(2\pi) - \frac{\phi_\tau}{2} - \frac{1}{2}\frac{\alpha_i^2}{\exp(\phi_\tau)}.
\end{equation*}
The components of $\nabla_{\btheta} \log p(\y_i, \alpha_i \mid \btheta) = (\nabla_{\bbeta} \log p(\y_i, \alpha_i \mid \btheta)^\top, \nabla_{\phi_\tau} \log p(\y_i, \alpha_i \mid \btheta))^\top$ are derived below:
\begin{align}
    \nabla_{\bbeta} \log p(\y_i, \alpha_i \mid \btheta) &= \nabla_{\bbeta} \log p(\y_i \mid \alpha_i, \bbeta) \quad \text{(since $\log p(\alpha_i \mid \btheta)$ does not depend on $\bbeta$)} \nonumber \\
    &= \sum_{j=1}^n y_{ij}[1 + \exp(-(\x_{ij}^\top \bbeta + \alpha_i))] \frac{\partial}{\partial \bbeta} \left( \frac{1}{1 + \exp(-(\x_{ij}^\top \bbeta + \alpha_i))}\right) \nonumber \\
    & \quad - \sum_{j=1}^n (1 - y_{ij}) \frac{1 + \exp(-(\x_{ij}^\top \bbeta + \alpha_i))}{\exp(-(\x_{ij}^\top \bbeta + \alpha_i))} \frac{\partial}{\partial \bbeta} \left( \frac{1}{1 + \exp(-(\x_{ij}^\top \bbeta + \alpha_i))}\right), \label{eq:logmm_grad_beta}
\end{align}
where
\begin{equation}
\label{eq:partial_pij}
    \frac{\partial}{\partial \bbeta} \left( \frac{1}{1 + \exp(-(\x_{ij}^\top \bbeta + \alpha_i))}\right) = \x_{ij} \frac{\exp(-(\x_{ij}^\top \bbeta + \alpha_i))}{[1 + \exp(-(\x_{ij}^\top \bbeta + \alpha_i))]^2}.
\end{equation}
Substituting~\eqref{eq:partial_pij} into~\eqref{eq:logmm_grad_beta} and reducing terms gives
\begin{align}
    \nabla_{\bbeta} \log p(\y_i, \alpha_i \mid \btheta)
    %
    &= \sum_{j=1}^n y_{ij} \x_{ij} \frac{\exp(-(\x_{ij}^\top \bbeta + \alpha_i))}{1 + \exp(-(\x_{ij}^\top \bbeta + \alpha_i))} - \sum_{j=1}^n (1 - y_{ij}) \x_{ij} \frac{1}{1 + \exp(-(\x_{ij}^\top \bbeta + \alpha_i))} \\
    &= \sum_{j=1}^n \x_{ij} \left[y_{ij} - \frac{1}{1 + \exp(-(\x_{ij}^\top \bbeta + \alpha_i))} \right].
\end{align}
The other component of $\nabla_{\btheta} \log p(\y_i, \alpha_i \mid \btheta)$ is  
\begin{align}
    \nabla_{\phi_\tau} \log p(\y_i, \alpha_i \mid \btheta) &= \nabla_{\phi_\tau} \log p(\alpha_i \mid \phi_\tau) \quad \text{(since $\log p(\y_i \mid \alpha_i, \bbeta)$ does not depend on $\phi_\tau$)} \nonumber \\
    &= -\frac{1}{2} + \frac{\alpha_i^2}{2\exp(\phi_\tau)}.
\end{align}
Evaluation of Louis' identity similarly requires
\begin{equation}
    \nabla^2_{\btheta} \log p(\y_i, \alpha_i \mid \btheta) = 
    \begin{bmatrix}
        \nabla^2_{\bbeta} \log p(\y_i, \alpha_i \mid \btheta) & \nabla_{\bbeta}\nabla_{\phi_\tau} \log p(\y_i, \alpha_i \mid \btheta) \\
        \nabla_{\bbeta}\nabla_{\phi_\tau} \log p(\y_i, \alpha_i \mid \btheta)^\top & \nabla^2_{\phi_\tau} \log p(\y_i, \alpha_i \mid \btheta)
    \end{bmatrix},
\end{equation}
the components of which are
\begin{align}
    \nabla^2_{\bbeta} \log p(\y_i, \alpha_i \mid \btheta) &= \sum_{j=1}^n \frac{\partial}{\partial \bbeta^\top} \left( \frac{\x_{ij}}{1 + \exp(-(\x_{ij}^\top \bbeta + \alpha_i))} \right) = - \sum_{j=1}^n \x_{ij} \x_{ij}^\top \frac{\exp(-(\x_{ij}^\top \bbeta + \alpha_i))}{\left[1 + \exp(-(\x_{ij}^\top \bbeta + \alpha_i)) \right]^2}, \\
    \nabla^2_{\phi_\tau} \log p(\y_i, \alpha_i \mid \btheta) &= -\frac{\alpha_i^2}{2 \exp(\phi_\tau)},  \\
    \nabla_{\bbeta}\nabla_{\phi_\tau} \log p(\y_i, \alpha_i \mid \btheta) &= \0.
\end{align}

\subsection{Derivation of gradient and Hessian expressions for the Poisson mixed model\label{derivativepoissonregression}}

In this section, we derive the gradient and Hessian for the Poisson model with bivariate random effects in Sect.~\ref{sec:poisson}. We note that the formula for the gradient in this section can be generalised to an arbitrary number of random effects. 

The parameters of interest in this model are the fixed effects, $\bbeta$, and the lower-diagonal elements of the Cholesky factor $\L$ of the random effect covariance matrix $\bSigma_\alpha$. We parameterise $\L$ as
\begin{equation}
    \L = \begin{bmatrix}
        \exp(\zeta_{11}) & 0 \\
        \zeta_{21} & \exp(\zeta_{22})  
    \end{bmatrix},
\end{equation}

and collect the parameters in a vector $\btheta = (\bbeta^\top, \bzeta)^\top$, where $\bzeta \equiv (\zeta_{11}, \zeta_{22}, \zeta_{21})^\top$. The likelihood function $p(\y_i \mid \btheta)$ is not available in closed form for this model, so the gradient and Hessian of the log likelihood with respect to the parameters $\btheta$ need to be approximated via Fisher's and Louis' identities. 

The evaluation of Fisher's identity requires the gradient of $\log p(\y_i, \balpha_i \mid \btheta)$, where
\begin{align}
    \log p(\y_i, \balpha_i \mid \btheta)
    &= \log p(\y_i \mid \balpha_i, \bbeta) + \log p(\balpha_i \mid \bzeta) \nonumber \\
    &= \sum_{j=1}^n [y_{ij} (\x_{ij}^\top \bbeta + \z_{ij}^\top \balpha_i) - \exp(\x_{ij}^\top \bbeta + \z_{ij}^\top \balpha_i)] \nonumber \\ 
    & \quad - \frac{r}{2}\log(2\pi) - \frac{1}{2} \log \abs{\bSigma_\alpha} -\frac{1}{2} \balpha_i^\top \bSigma_\alpha^{-1} \balpha_i,
\end{align}
for $i = 1, \dots, N, \, j = 1, \dots, n$, and where $r$ is the number of random effects. The gradient is $${\nabla_{\btheta} \log p(\y_i, \balpha_i \mid \btheta) = (\nabla_{\bbeta} \log p(\y_i, \balpha_i \mid \btheta), \nabla_{\bzeta} \log p(\y_i, \balpha_i \mid \btheta))^\top}.$$ As $\log p(\balpha_i \mid \bzeta)$ does not depend on $\bbeta$, the gradient with respect to $\bbeta$ is simply
\begin{equation}
    \nabla_{\bbeta} \log p(\y_i, \balpha_i \mid \btheta) = \nabla_{\bbeta} \log p(\y_i \mid \balpha_i, \bbeta) = \sum_{j=1}^n [y_{ij} - \exp(\x_{ij}^\top \bbeta + \z_{ij}^\top \balpha_i)] \x_{ij}.
\end{equation}
Similarly, as $\log p(\y_i \mid \balpha_i, \bbeta)$ does not depend on $\bzeta$, the gradient with respect to $\bzeta$ reduces to 
\begin{align*}
    \nabla_{\bzeta} \log p(\y_i, \balpha_i \mid \btheta) &= \nabla_{\bzeta} \log p(\balpha_i \mid \bzeta) \nonumber \\
    &= \Tr(\A^\top \nabla_{\zeta_{kl}} \L), \quad k = 1, 2, l < k,
\end{align*}
where $\A = -\L^{-\top} + \L^{-\top} \L^{-1} \balpha_i \balpha_i^\top \L^{-\top}$, with $\M^{-\top}$ denoting the inverse of the transpose of a matrix $\M$, and  
\begin{equation}
    \nabla_{\zeta_{kl}} \L =
    \begin{cases} 
    \J^*_{kk} & \text{if } k = l, \\
    \J_{kl} & \text{otherwise}, 
    \end{cases} 
\end{equation}
for $k = 1, 2, \, l < k$, where $\J_{kl}$ is a $2 \times 2$ matrix that has 1 in the $(k,l)$th element and 0 elsewhere, and $\J^*_{kk}$ is a $2 \times 2$ matrix that has $\exp(\zeta_{kk})$ in the $(k,k)$th element and 0 elsewhere.

The Hessian for the subject-specific log likelihood is
\begin{align*}
    \nabla^2_{\btheta} \log p(\y_i, \balpha_i \mid \btheta) &= 
    \begin{bmatrix}
        \nabla^2_{\bbeta} \log p(\y_i, \balpha_i \mid \btheta) & \nabla_{\bzeta} \nabla_{\bbeta} \log p(\y_i, \balpha_i \mid \btheta)^\top \\
        \nabla_{\bzeta} \nabla_{\bbeta} \log p(\y_i, \balpha_i \mid \btheta) & \nabla^2_{\bzeta} \log p(\y_i, \balpha_i \mid \btheta) \\
    \end{bmatrix} \\
    &= 
    \begin{bmatrix}
        \nabla^2_{\bbeta} \log p(\y_i, \balpha_i \mid \btheta) & \0_{4 \times 3} \\
        \0_{3 \times 4} & \nabla^2_{\bzeta} \log p(\y_i, \balpha_i \mid \btheta) \\
    \end{bmatrix},
\end{align*}
where $\0_{m \times n}$ is a matrix of size $m \times n$. Here the cross-derivative $\nabla_{\bzeta} \nabla_{\bbeta} \log p(\y_i, \balpha_i \mid \btheta)$ is zero because $\nabla_{\bbeta} \log p(\y_i, \balpha_i \mid \btheta)$ does not depend on $\bzeta$. The remaining components of the Hessian matrix are
\begin{equation}
    \nabla^2_{\bbeta} \log p(\y_i, \balpha_i \mid \btheta) = \sum_{j=1}^n [y_{ij} - \exp(\x_{ij}^\top \bbeta + \z_{ij}^\top \balpha_i)] \x_{ij} \x_{ij}^\top,
\end{equation}
and 
\begin{equation*}
    \nabla^2_{\bzeta} \log p(\y_i, \balpha_i \mid \btheta) = 
    \begin{bmatrix}
        \nabla^2_{\zeta_{11}} \log p(\y_i, \balpha_i \mid \btheta) & \nabla_{\zeta_{22}} \nabla_{\zeta_{11}}\log p(\y_i, \balpha_i \mid \btheta)^\top & \nabla_{\zeta_{21}} \nabla_{\zeta_{11}}\log p(\y_i, \balpha_i \mid \btheta)^\top \\
        \nabla_{\zeta_{22}} \nabla_{\zeta_{11}} \log p(\y_i, \balpha_i \mid \btheta) & \nabla^2_{\zeta_{22}} \log p(\y_i, \balpha_i \mid \btheta)^\top & \nabla_{\zeta_{21}} \nabla_{\zeta_{22}}\log p(\y_i, \balpha_i \mid \btheta)^\top \\
        \nabla_{\zeta_{21}} \nabla_{\zeta_{11}} \log p(\y_i, \balpha_i \mid \btheta) & \nabla_{\zeta_{21}} \nabla_{\zeta_{22}}\log p(\y_i, \balpha_i \mid \btheta)^\top & \nabla^2_{\zeta_{21}} \log p(\y_i, \balpha_i \mid \btheta)^\top \\
    \end{bmatrix}
\end{equation*}
where
\begin{equation}
    \nabla^2_{\zeta_{kl}} \log p(\y_i, \balpha_i \mid \btheta) = \Tr( (\nabla_{\zeta_{kl}} \A)^\top \nabla_{\zeta_{kl}} \L + \A^\top \nabla^2_{\zeta_{kl}} \L), \quad  k = 1, 2, \, l < k.
\end{equation}
Separately,
\begin{equation}
    \nabla^2_{\zeta_{kl}} \L = 
    \begin{cases}
        \J^*_{kk}, & \text{if } k = l \\
        \0_{2 \times 2}, & \text{otherwise},
    \end{cases}
\end{equation}
and
\begin{align}
    \nabla_{\zeta_{kl}} \A 
    &= \nabla_{\zeta_{kl}} (-\L^{-\top} + \L^{-\top} \L^{-1} \balpha_i \balpha_i^\top \L^{-\top}) \nonumber \\
    &= - \nabla_{\zeta_{kl}} \L^{-\top} + (\nabla_{\zeta_{kl}} \L^{-\top}) \B + \L^{-\top} \nabla_{\zeta_{kl}} \B,
\end{align}
where $\B = \L^{-1} \balpha_i \balpha_i^\top \L^{-\top}$, and $\nabla_{\zeta_{kl}} \B = (\nabla_{\zeta_{kl}} \L^{-1}) \balpha_i \balpha_i^\top \L^{-\top} + \L^{-1} \balpha_i \balpha_i^\top (\nabla_{\zeta_{kl}} \L^{-\top})$, for $k = 1, 2, \, l < k$.
We derive the terms $\nabla_{\zeta_{kl}} \L^{-\top}$ individually:
\begin{align*}
    \nabla_{\zeta_{11}} \L^{-\top} &= 
    \begin{bmatrix}
    - \exp(-\zeta_{11}) & 0 \\
    \zeta_{21} \exp(-\zeta_{11} - \zeta_{22}) & 0
    \end{bmatrix}, \\
    \nabla_{\zeta_{22}} \L^{-\top} &= 
    \begin{bmatrix}
    0 & 0 \\
    \zeta_{21} \exp(-\zeta_{11} - \zeta_{22}) & - \exp(-\zeta_{22}) \\
    \end{bmatrix}, \\
    \nabla_{\zeta_{21}} \L^{-\top} &= 
    \begin{bmatrix}
    0 & 0 \\
    -\exp(-\zeta_{11} - \zeta_{22}) & 0 \\
    \end{bmatrix}.
\end{align*}


    \clearpage
\section{Variance of the R-VGAL posterior densities for various Monte Carlo sample sizes}
\label{sec:var_test_S_Salpha}

In this section, we study the effects of increasing the number of samples $S$, for estimating the expectations in the variational mean and precision matrix updates, and $S_\alpha$, for estimating the gradient of the log-likelihood
on the R-VGAL posterior estimates. 
The results are obtained using the simulated logistic data in Sect.~\ref{sec:logmm} and the Polypharmacy dataset in Sect.~\ref{sec:realdata} of the main paper. Similar results are obtained for other datasets. For all simulations in this section, we use the damped R-VGAL algorithm in Sect.~\ref{sec:damping} with $n_{damp} = 10$ observations and $K = 4$ damping steps per observation.

\subsection{Simulated logistic data \label{simulatedlogisticdatasamplesizes}}
\begin{figure}[ht]
    \centering
    \includegraphics[width = \linewidth]{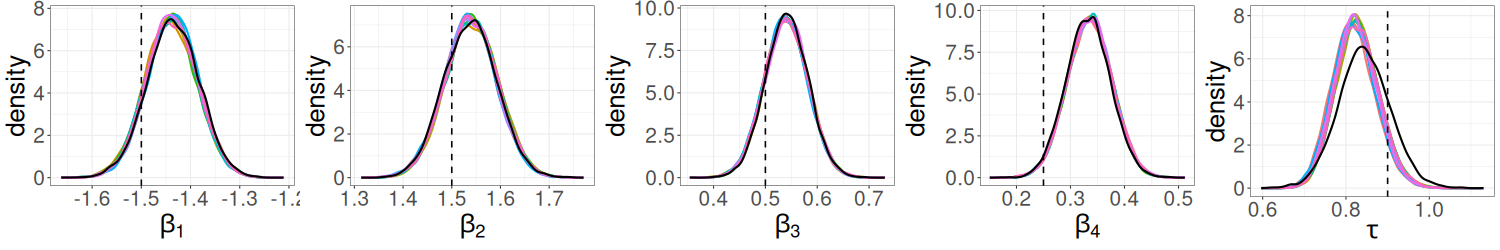}
    \caption{R-VGAL posterior distributions from 10 runs on simulated logistic data are plotted in colour. Damping is done on the first 10 observations, and the Monte Carlo sample sizes are $S = 50$ and $S_\alpha = 50$. HMC posterior distributions are plotted in black for comparison.}
    \label{fig:logistic_var_test_temper10_S50_Sa50}
\end{figure}

\begin{figure}[ht]
    \centering
    \includegraphics[width = \linewidth]{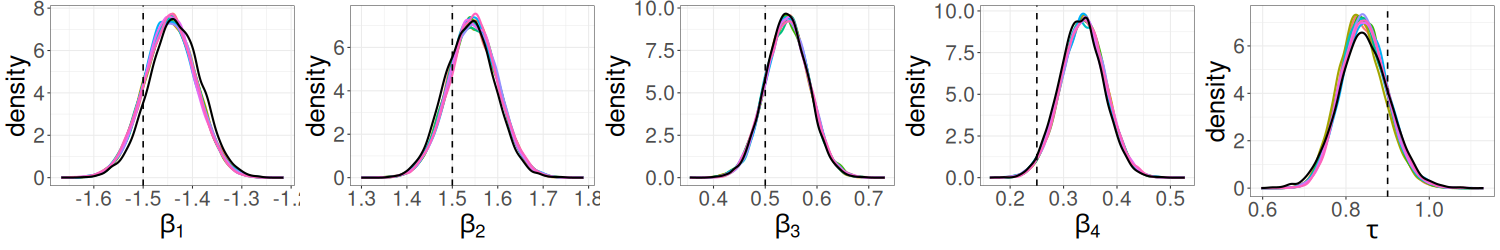}
    \caption{R-VGAL posterior distributions from 10 runs on simulated logistic data are plotted in colour. Damping is done on the first 10 observations, and the Monte Carlo sample sizes are $S = 100$ and $S_\alpha = 100$. HMC posterior distributions are plotted in black for comparison.}
    \label{fig:logistic_var_test_temper10_S100_Sa100}
\end{figure}

\begin{figure}[ht]
    \centering
    \includegraphics[width = \linewidth]{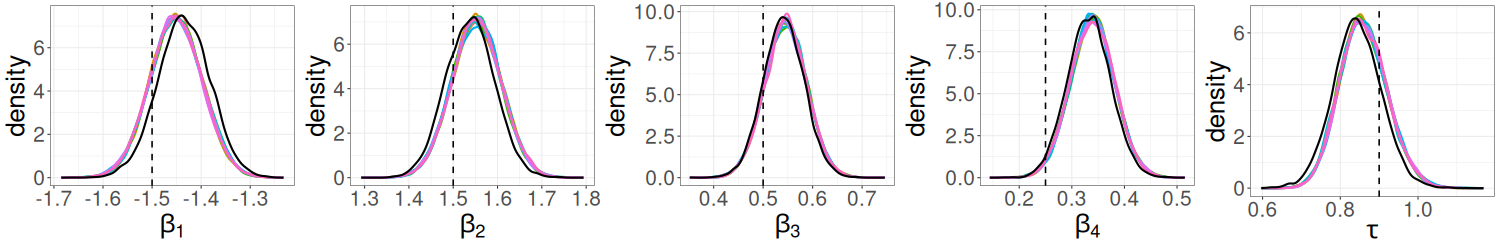}
    \caption{R-VGAL posterior distributions from 10 runs on simulated logistic data are plotted in colour. Damping is done on the first 10 observations, and the Monte Carlo sample sizes are $S = 500$ and $S_\alpha = 500$. HMC posterior distributions are plotted in black for comparison.}
    \label{fig:logistic_var_test_temper10_S500_Sa500}
\end{figure}

\begin{figure}[ht]
    \centering
    \includegraphics[width = \linewidth]{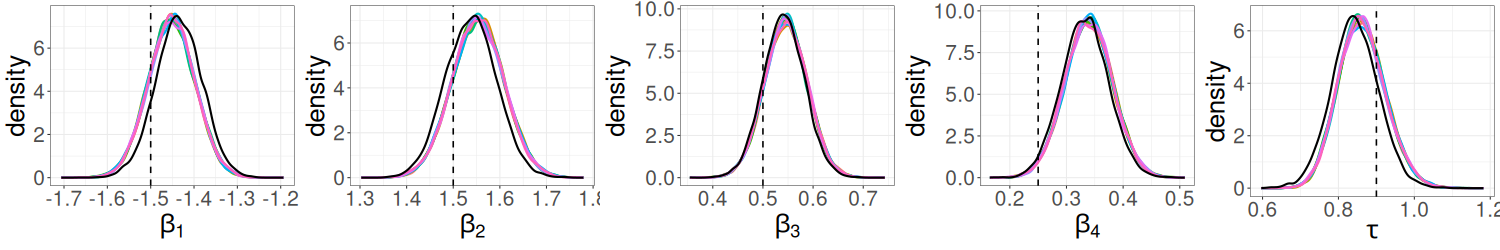}
    \caption{R-VGAL posterior distributions from 10 runs on simulated logistic data are plotted in colour. Damping is done on the first 10 observations, and the Monte Carlo sample sizes are $S = 1000$ and $S_\alpha = 1000$. HMC posterior distributions are plotted in black for comparison.}
    \label{fig:logistic_var_test_temper10_S1000_Sa1000}
\end{figure}

The simulated data used in this section is the same as that used in Sect.~\ref{sec:logmm} of the main paper. The same values for $S$ and $S_\alpha$ are used and they are taken from the set $\{50, 100, 500, 1000\}$. For each pair of $S$ and $S_\alpha$ values, we independently run R-VGAL 10 times on the simulated dataset, and plot the posterior densities from all 10 runs for each parameter. These posterior densities are shown in Figures~\ref{fig:logistic_var_test_temper10_S50_Sa50} to~\ref{fig:logistic_var_test_temper10_S1000_Sa1000}. For comparison, the HMC posterior distributions (from a single run, with 2 chains and 20000 total posterior samples after burn-in) are also plotted in each figure.

As the Monte Carlo sample sizes increase, the R-VGAL posterior density estimates get closer and closer to each other, which shows that increasing the values of $S$ and $S_\alpha$ helps reduce the variability of the R-VGAL posterior estimates across multiple independent runs.

\subsection{Polypharmacy dataset}
\begin{figure}[ht]
    \centering
    \includegraphics[width = 0.8\linewidth]{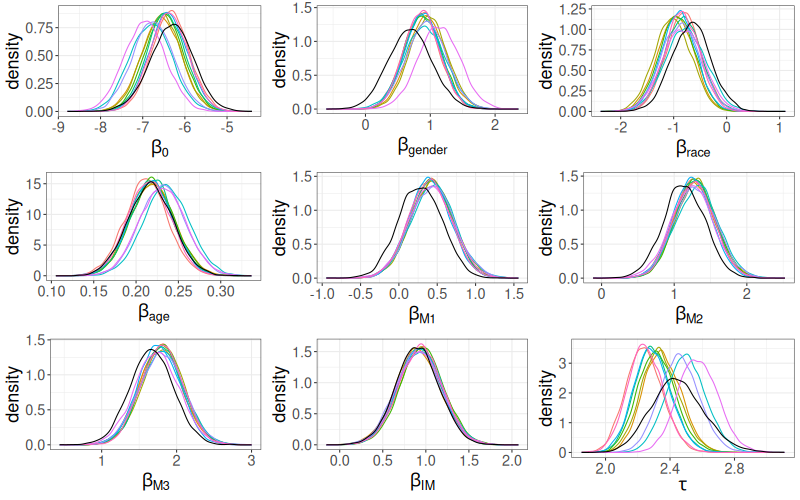}
    \caption{R-VGAL posterior distributions from 10 runs on the Polypharmacy dataset are plotted in colour. Damping is done on the first 10 observations, and the Monte Carlo sample sizes are $S = 50$ and $S_\alpha = 50$. HMC posterior distributions are plotted in black for comparison.}
    \label{fig:var_test_temper10_S50_Sa50_20230327_1}
\end{figure}

\begin{figure}[ht]
    \centering
    \includegraphics[width = 0.8\linewidth]{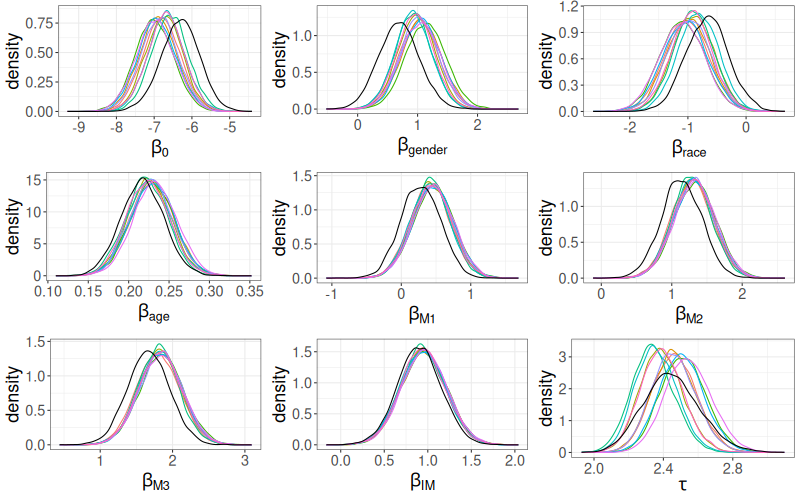}
    \caption{R-VGAL posterior distributions from 10 runs on the Polypharmacy dataset are plotted in colour. Damping is done on the first 10 observations, and the Monte Carlo sample sizes are $S = 100$ and $S_\alpha = 100$. HMC posterior distributions are plotted in black for comparison.}
    \label{fig:var_test_temper10_S100_Sa100_20230327_1}
\end{figure}

\begin{figure}[ht]
    \centering
    \includegraphics[width = 0.8\linewidth]{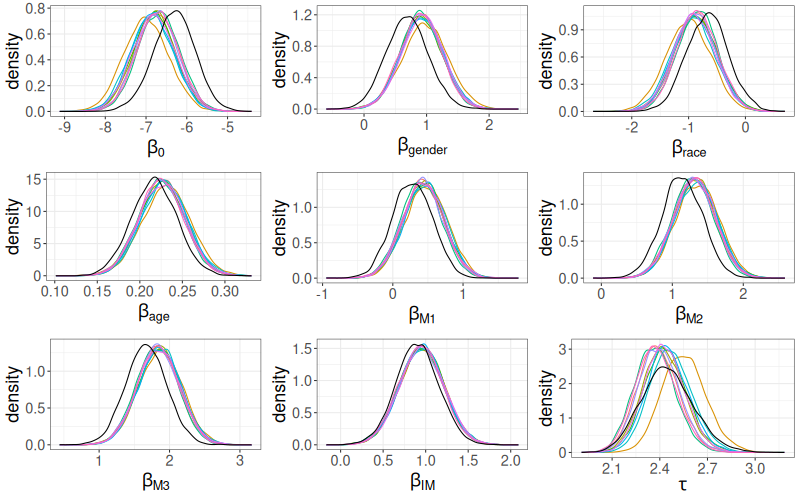}
    \caption{R-VGAL posterior distributions from 10 runs on the Polypharmacy dataset are plotted in colour. Damping is done on the first 10 observations, and the Monte Carlo sample sizes are $S = 500$ and $S_\alpha = 500$. HMC posterior distributions are plotted in black for comparison.}
    \label{fig:var_test_temper10_S500_Sa500_20230327_1}
\end{figure}

\begin{figure}[ht]
    \centering
    \includegraphics[width = 0.8\linewidth]{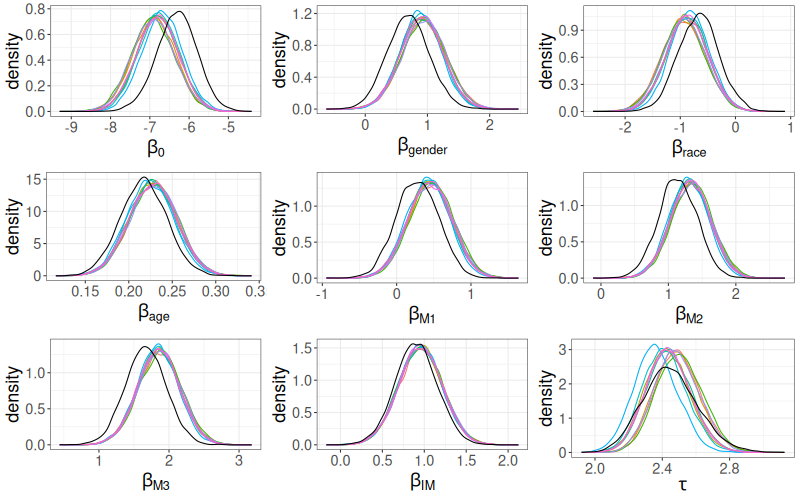}
    \caption{R-VGAL posterior distributions from 10 runs on the Polypharmacy dataset are plotted in colour. Damping is done on the first 10 observations, and the Monte Carlo sample sizes are $S = 1000$ and $S_\alpha = 1000$. HMC posterior distributions are plotted in black for comparison.}
    \label{fig:var_test_temper10_S1000_Sa1000_20230327_1}
\end{figure}

The Polypharmacy dataset used in this section is the same as that used in Sect.~\ref{sec:realdata} of the main paper. Similar to Sect. \ref{simulatedlogisticdatasamplesizes}, the same values for $S$ and $S_\alpha$ are used and they are taken from the set $\{50, 100, 500, 1000\}$. We independently run R-VGAL 10 times on the Polypharmacy dataset for each pair of $S$ and $S_\alpha$ values, and plot the posterior densities from all 10 runs for each parameter.  These posteriors are shown in Figures~\ref{fig:var_test_temper10_S50_Sa50_20230327_1} to~\ref{fig:var_test_temper10_S1000_Sa1000_20230327_1}. For comparison, the HMC posterior distributions (from a single run, with 2 chains and 20000 total posterior samples after burn-in) are also plotted.

\clearpage
As with the previous example, the results in this example also show that increasing the values of $S$ and $S_\alpha$ reduces the variability of the R-VGAL posterior estimates across multiple runs. This phenomenon is particularly pronounced for the random effect standard deviation $\tau$. Suitable values for $S$ and $S_\alpha$ are likely to be application-dependent. However, from our studies, we conclude that $S$ and $S_\alpha$ need to be at least 100 for the Monte Carlo sample sizes not to have a substantial effect on the final estimates.
    \section{Robustness check of the R-VGAL algorithm
\label{sec:sensitivityRVGAL}}

In this section, we use the Polypharmacy dataset in Sect.~\ref{sec:realdata} to check the robustness of the R-VGAL algorithm given different orderings of the data. The simulations in this section show that R-VGAL can be unstable while traversing the first few observations, which makes it sensitive to the ordering of observations. This instability can, however, be alleviated with the damped R-VGAL algorithm, as described in Sect.~\ref{sec:damping} of the main paper.

\begin{figure}[!ht]
    \centering
    \includegraphics[width = 0.9\linewidth]{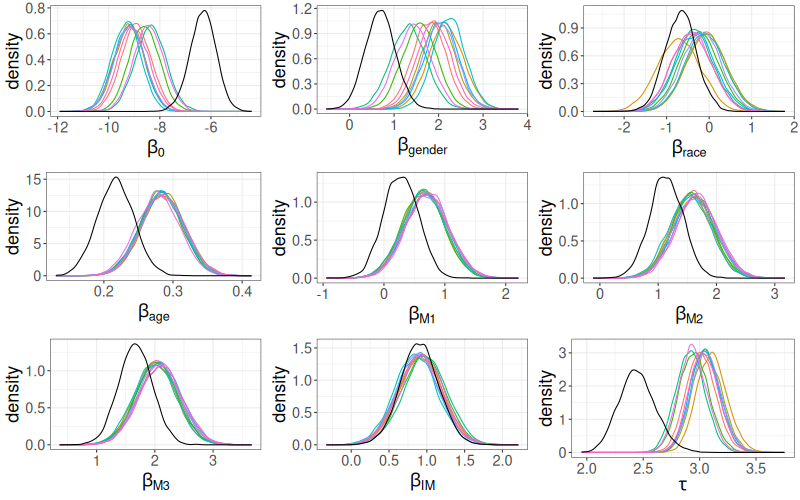}
    \caption{R-VGAL posterior distributions from 10 independent runs using the original ordering of the data are plotted in different colours. The Monte Carlo sample sizes are $S = 100, S_\alpha = 100$. HMC posterior distributions are plotted in black for comparison.}
    \label{fig:var_test_S100_Sa100}
\end{figure}

\begin{figure}[ht]
    \centering
    \includegraphics[width = 0.9\linewidth]{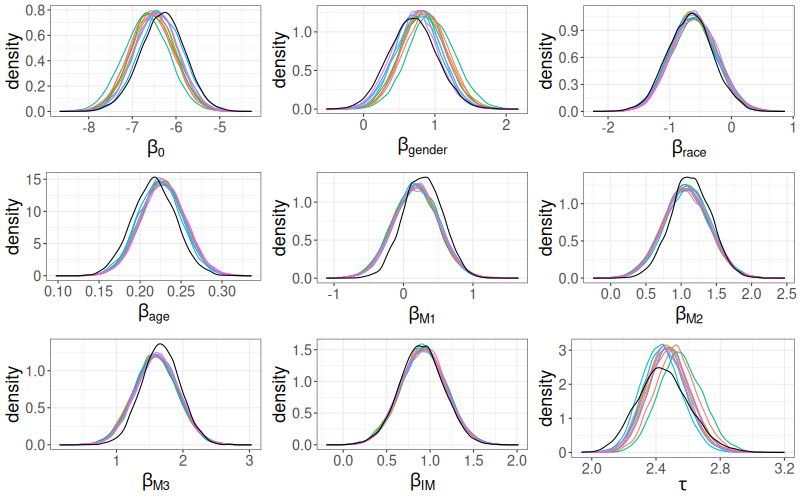}
    \caption{R-VGAL posterior distributions from 10 independent runs using the random ordering of the data are plotted in different colours. The Monte Carlo sample sizes are $S = 100$, and $S_\alpha = 100$. HMC posterior distributions are plotted in black for comparison.}
    \label{fig:var_test_seed2023_S100_Sa100}
\end{figure}

Figures \ref{fig:var_test_S100_Sa100} and \ref{fig:var_test_seed2023_S100_Sa100} show the R-VGAL posterior density estimates from 10 independent runs using the original ordering of the data and a random ordering of the data, respectively. In both simulations, the number of Monte Carlo samples $S$ to estimate the expectation with respect to $q_{i-1}(\btheta)$ and the number of samples $S_\alpha$ to estimate the gradients/Hessians are fixed to $100$. In both figures, the HMC posterior densities for each parameter are plotted in black for comparison. The figures show that the R-VGAL estimates are quite far away from those of HMC estimates when using the original ordering of the data, while the R-VGAL estimates are reasonably close to those of HMC when using the random ordering of the data. This suggests that the R-VGAL estimates are not robust with respect to the ordering of the data.

\begin{figure}[!ht]
    \centering
    \includegraphics[width = 0.9\linewidth]{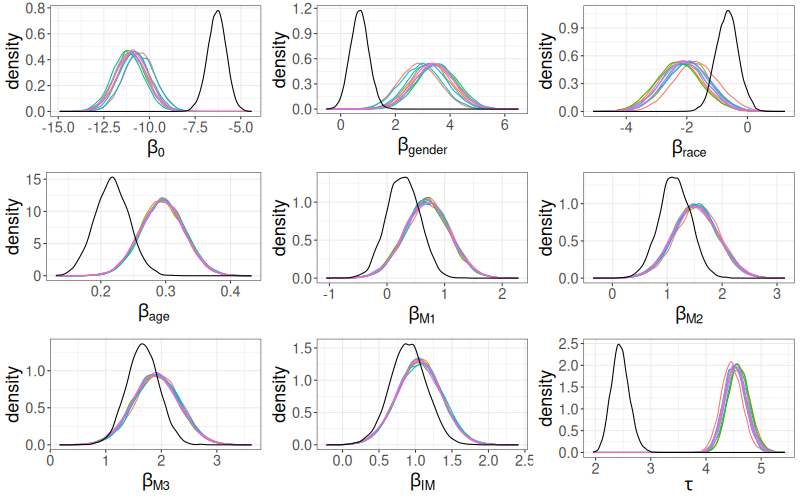}
    \caption{R-VGAL posterior distributions from 10 independent runs using the original ordering of the data are plotted in colour. The Monte Carlo sample sizes are $S = 1000, S_\alpha = 1000$. HMC posterior distributions are plotted in black for comparison.}
    \label{fig:var_test_S1000_Sa1000}
\end{figure}

\begin{figure}[!ht]
    \centering
    \includegraphics[width = 0.9\linewidth]{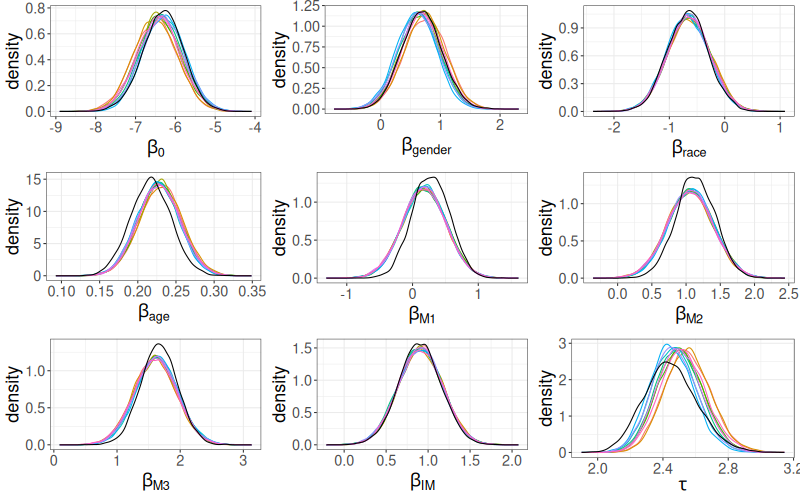}
    \caption{R-VGAL posterior distributions from 10 independent runs using the random ordering of the data are plotted in colour. The Monte Carlo sample sizes are $S = 1000, S_\alpha = 1000$. HMC posterior distributions are plotted in black for comparison.}
    \label{fig:var_test_seed2023_S1000_Sa1000}
\end{figure}

To confirm that the source of variability in the R-VGAL estimates is from different data orderings and not from the low number of Monte Carlo samples, we increase the number of Monte Carlo samples $S$ and $S_\alpha$. Figures \ref{fig:var_test_S1000_Sa1000} and \ref{fig:var_test_seed2023_S1000_Sa1000} show the R-VGAL posterior density estimates from 10 independent runs using the original ordering of the data and a random ordering of the data, respectively, with the Monte Carlo sample sizes set to $S = S_{\alpha} = 1000$. The posterior densities for each parameter are different for the two orderings; for instance, with the original ordering, the posterior of $\tau$ is centred around 4.5, while with the random ordering, the posterior of $\tau$ is centred around 2.4.

\begin{figure}[!ht]
    \centering
    \includegraphics[width = 0.9\linewidth]{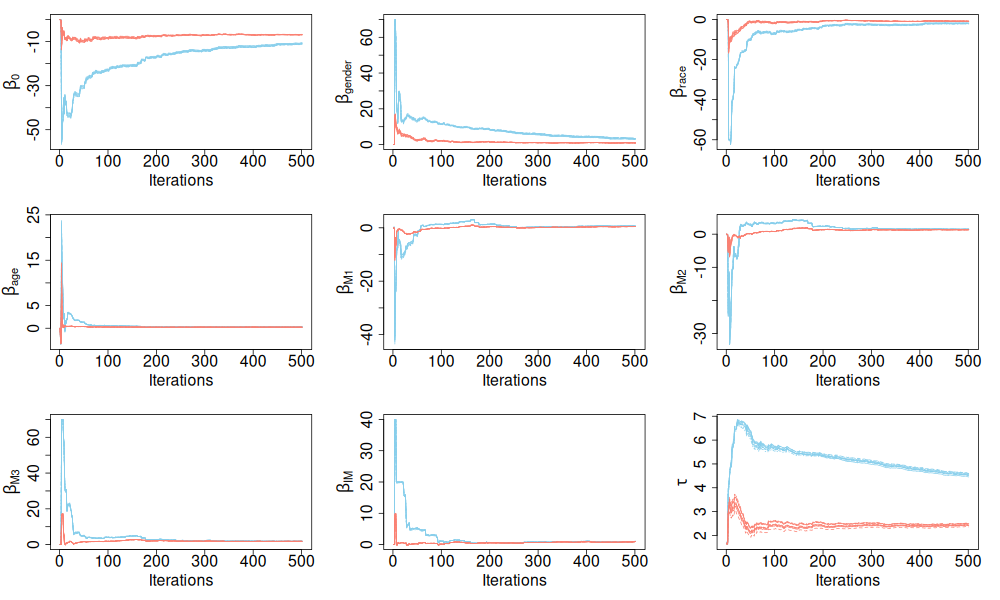}
    \caption{Trajectories of the variational mean without damping (in blue) and with damping (in red) for each parameter across 10 independent runs of R-VGAL on the original ordering of the data. The Monte Carlo sample sizes are $S = 1000$ and $S_\alpha = 1000$.}
    \label{fig:trajectories_S1000_Sa1000}
\end{figure}

\begin{figure}[!ht]
    \centering
    \includegraphics[width = 0.9\linewidth]{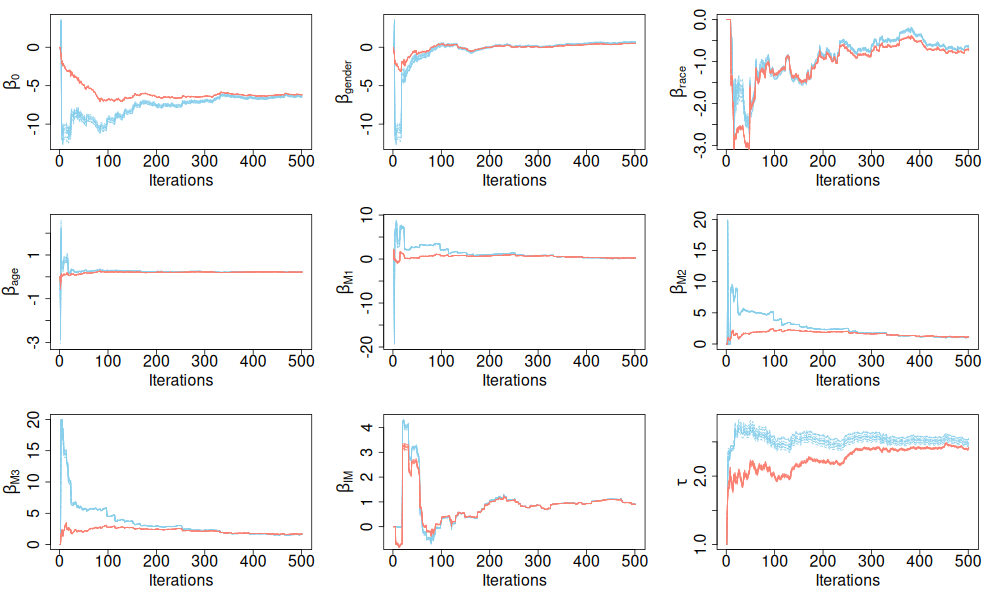}
    \caption{Trajectories of the variational mean without damping (in blue) and with damping (in red) for each parameter across 10 independent runs of R-VGAL on the random ordering of the data. The Monte Carlo sample sizes are $S = 1000$ and $S_\alpha = 1000$.}
    \label{fig:trajectories_seed2023_S1000_Sa1000}
\end{figure}

A plot of the trajectory of the variational mean across R-VGAL iterations reveals that R-VGAL is unstable during the first few iterations. The blue lines in Figures~\ref{fig:trajectories_S1000_Sa1000} and~\ref{fig:trajectories_seed2023_S1000_Sa1000} show the trajectories of the variational mean for each of the parameters across 10 independent runs of the R-VGAL algorithm, on the original ordering and on a random ordering of the data, respectively. The initial trajectories of the fixed effect parameters in Figure \ref{fig:trajectories_S1000_Sa1000} vary significantly (for example, between -50 and 0 for the intercept $\beta_0$), and the trajectory of $\tau$ is dragged up to nearly 7 before progressively dropping towards 4. This potentially contributes to the biased posterior mean of $\tau$. In Figure~\ref{fig:trajectories_seed2023_S1000_Sa1000}, where the data were randomly reordered, the trajectories of the parameters are less variable initially, which then allows the variational mean to converge towards the true values more rapidly. This shows that the R-VGAL algorithm is unstable while traversing the first few observations, making the algorithm sensitive to the ordering of the data.

\begin{figure}[!t]
    \centering
    \includegraphics[width = 0.9\linewidth]{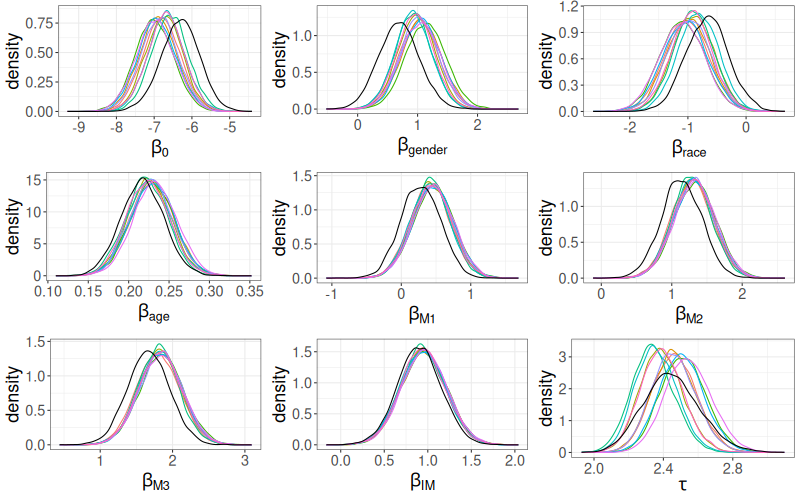}
    \caption{R-VGAL posterior distributions from 10 independent runs using the original ordering of the data are plotted in colour. Damping is done on the first 10 observations. The Monte Carlo sample sizes are $S = 100, S_\alpha = 100$. HMC posterior distributions are plotted in black for comparison.}
    \label{fig:var_test_temper10_S100_Sa100}
\end{figure}

\begin{figure}[!ht]
    \centering
    \includegraphics[width = 0.9\linewidth]{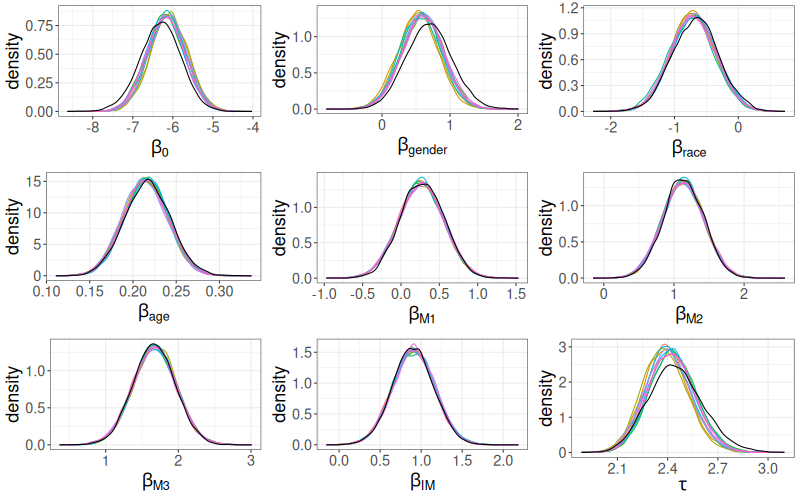}
    \caption{R-VGAL posterior distributions from 10 independent runs using a random reordering of the data are plotted in colour. Damping is done on the first 10 observations. The Monte Carlo sample sizes are $S = 100, S_\alpha = 100$. HMC posterior distributions are plotted in black for comparison.}
    \label{fig:var_test_temper10_seed2023_S100_Sa100}
\end{figure}

We propose a damping approach (in Sect.~\ref{sec:damping} of the main paper) to make the R-VGAL algorithm more robust. By damping the first few observations, the R-VGAL posterior estimates become much more consistent across different data orderings. 
Figures \ref{fig:var_test_temper10_S100_Sa100} and \ref{fig:var_test_temper10_seed2023_S100_Sa100} show the posterior density estimates from 10 independent runs of the R-VGAL algorithm using the original ordering of the data and a random ordering of the data, respectively, with damping done on the first 10 observations, and with Monte Carlo sample sizes $S = S_\alpha = 100$. These figures show that the posterior density estimates of R-VGAL with damping are consistent across two different orderings of the data, and also consistent with those obtained from HMC. 

\begin{figure}[!t]
    \centering
    \includegraphics[width = 0.9\linewidth]{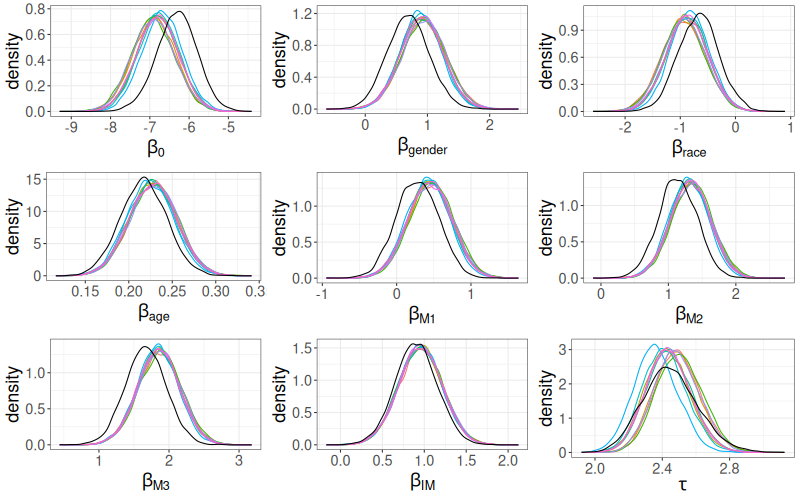}
    \caption{R-VGAL posterior distributions from 10 independent runs using the original ordering of the data are plotted in colour. Damping is done on the first 10 observations. The Monte Carlo sample sizes are $S = 1000, S_\alpha = 1000$. HMC posterior distributions are plotted in black for comparison.}
    \label{fig:var_test_temper10_S1000_Sa1000}
\end{figure}

\begin{figure}[!t]
    \centering
    \includegraphics[width = 0.9\linewidth]{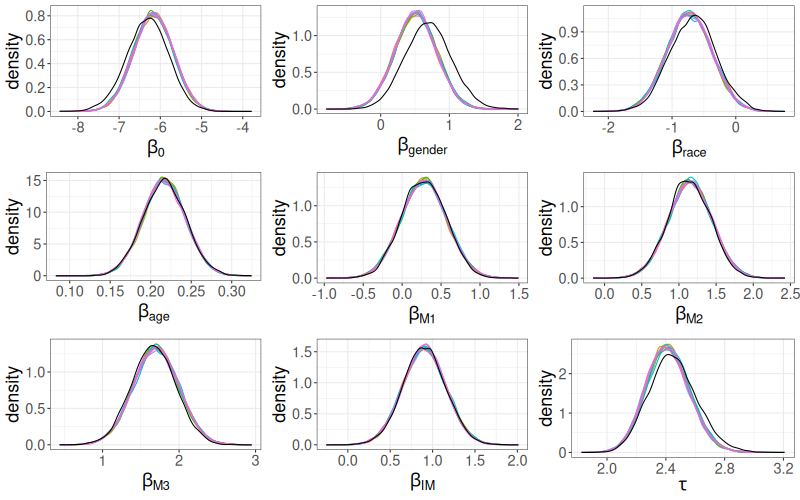}
    \caption{R-VGAL posterior distributions from 10 independent runs using a random reordering of the data are plotted in colour. Damping is done on the first 10 observations. The Monte Carlo sample sizes are $S = 1000, S_\alpha = 1000$. HMC posterior distributions are plotted in black for comparison. }
    \label{fig:var_test_temper10_seed2023_S1000_Sa1000}
\end{figure}

Figures~\ref{fig:var_test_temper10_S1000_Sa1000} and~\ref{fig:var_test_temper10_seed2023_S1000_Sa1000} display the R-VGAL posterior densities using the original ordering and the random ordering of the data, respectively, with damping done on the first 10 observations, and Monte Carlo sample sizes increased to $S = S_\alpha = 1000$. There is now very little difference between the posterior densities using the original and the random ordering of the dataset. 
The red lines in Figures~\ref{fig:trajectories_S1000_Sa1000} and~\ref{fig:trajectories_seed2023_S1000_Sa1000} show the parameter trajectories obtained from damped R-VGAL, on the original ordering and the random ordering of the data, respectively. The trajectories with damping (plotted in red) are much more stable than those without damping (plotted in blue), especially during the first few iterations. These figures suggest that damping is effective in reducing the variability of R-VGAL estimates while traversing the first few observations and increases the algorithm's robustness to different data orderings. Other random orderings of the data give similar results.

    \clearpage
\section{Repeated simulations \label{sec:multi_sims}}
In this section, for each of the linear, logistic and Poisson mixed models, we simulate 100 datasets with the same parameter settings. We run R-VGAL and HMC on each of the 100 datasets, and compare their posterior density estimates.

\subsection{Repeated simulations from the linear mixed model \label{sec:multi_sims_linear}}

\begin{figure}
    \centering
    \includegraphics[width = \linewidth]{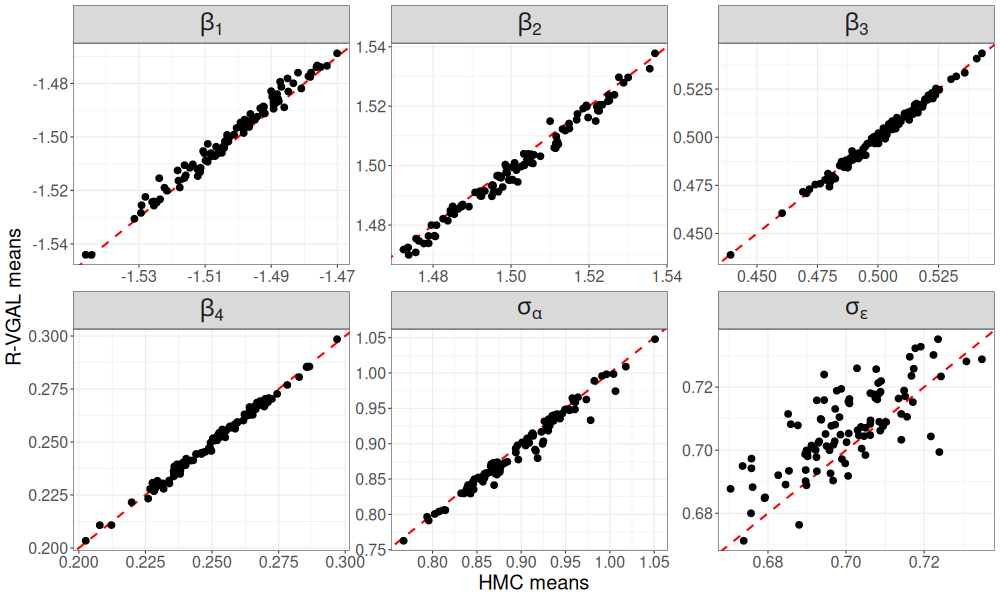}
    \caption{Plot of the R-VGAL posterior means against those from HMC for 100 datasets simulated from the linear mixed model.}
    \label{fig:linear_multi_sim_means_20230329}
\end{figure}

\begin{figure}
    \centering
    \includegraphics[width = \linewidth]{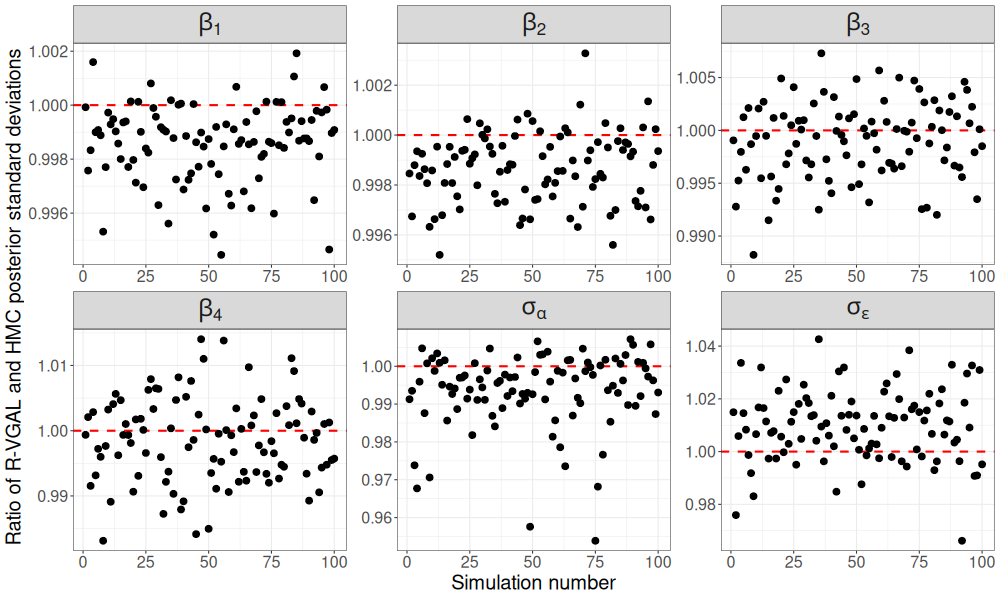}
    \caption{Plot of the ratio between the R-VGAL posterior standard deviations and those from HMC for 100 datasets simulated from the linear mixed model.}
    \label{fig:linear_multi_sim_sds_20230329}
\end{figure}

\begin{figure}
    \centering
    \includegraphics[width = \linewidth]{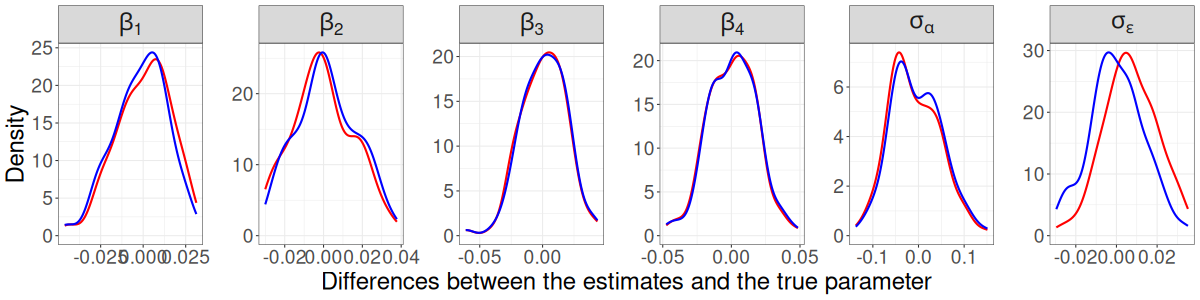}
    \caption{Comparison of the distribution of the differences between the R-VGAL posterior means and the true parameters (in red), against the distribution of the differences between the HMC posterior means and the true parameters (in blue), for 100 datasets simulated from the linear mixed model.}
    \label{fig:linear_multi_sim_difference_dens_20230329}
\end{figure}

The synthetic datasets in this section are simulated according to the number of observations and parameter values detailed in Sect.~\ref{sec:lmm}. Figure~\ref{fig:linear_multi_sim_means_20230329} plots the R-VGAL posterior means against those from HMC for each of the 100 simulated datasets. The red dotted diagonal line marks the ``ideal" scenario where the posterior means from the two methods are equal. Figure~\ref{fig:linear_multi_sim_sds_20230329} plots the ratio between the R-VGAL and HMC posterior standard deviations, with the dotted horizontal line marking the ideal ratio of one. Figure~\ref{fig:linear_multi_sim_difference_dens_20230329} compares the distribution of the differences between the R-VGAL means and the true parameter values to the distribution of the differences between the HMC means and the true parameter values. {We see that posterior means and standard deviations from the two methods are very similar across the 100 replicated datasets.}

\clearpage
\subsection{Repeated simulations from the logistic mixed model \label{sec:multi_sims_logistic}}
\begin{figure}
    \centering
    \includegraphics[width = \linewidth]{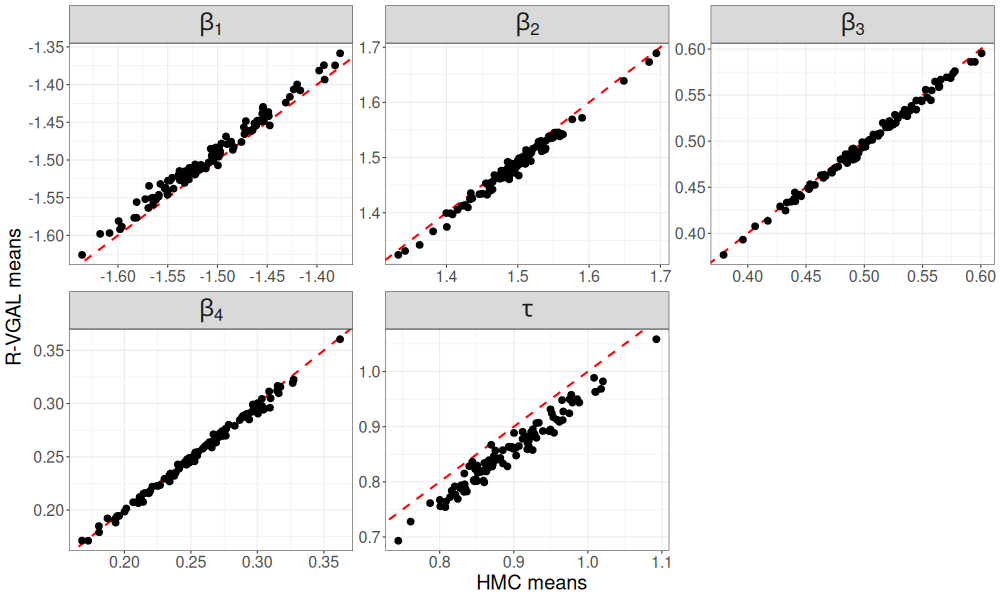}
    \caption{Plot of the R-VGAL posterior means against those from HMC for 100 datasets simulated from the logistic mixed model.}
    \label{fig:logistic_multi_sim_means_20231017}
\end{figure}

\begin{figure}
    \centering
    \includegraphics[width = \linewidth]{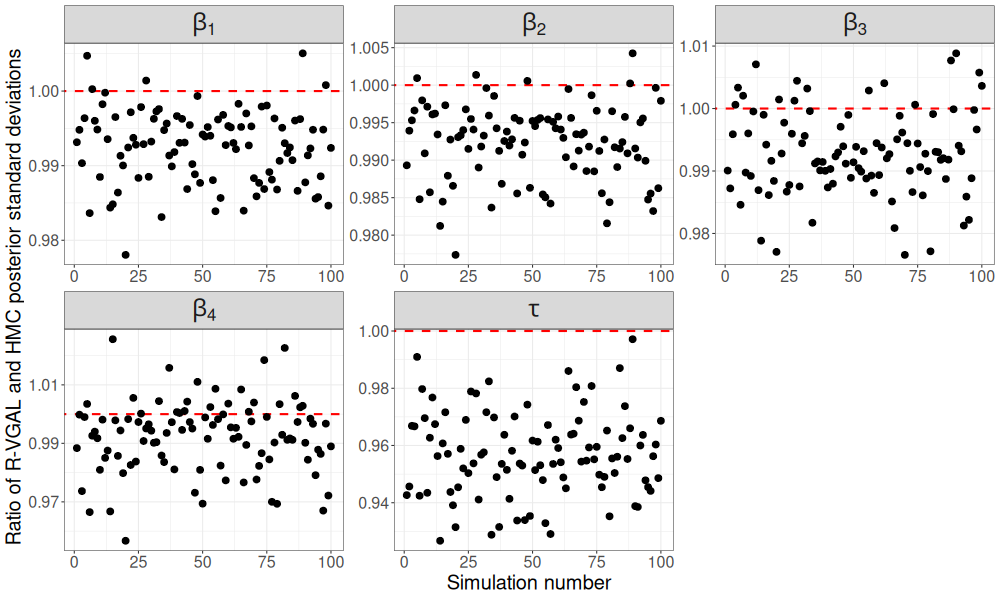}
    \caption{Plot of the ratio between the R-VGAL posterior standard deviations and those from HMC for 100 datasets simulated from the logistic mixed model.}
    \label{fig:logistic_multi_sim_sds_20231017}
\end{figure}

\begin{figure}
    \centering
    \includegraphics[width = \linewidth]{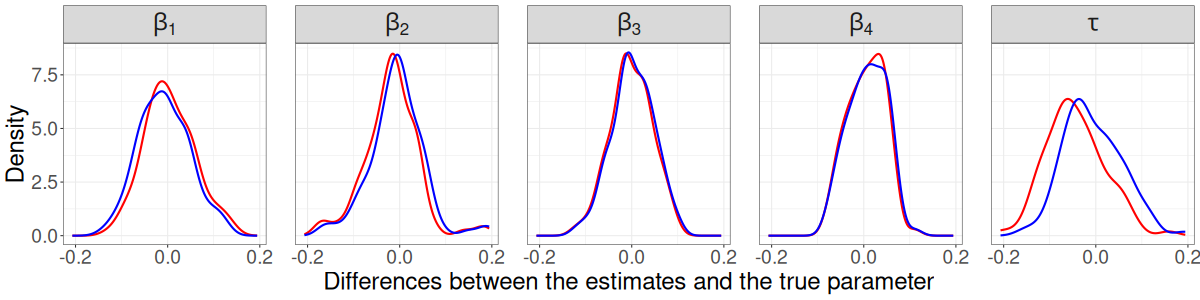}
    \caption{Comparison of the distribution of the differences between the R-VGAL posterior means and the true parameters (in red), against the distribution of the differences between the HMC posterior means and the true parameters (in blue), for 100 datasets simulated from the logistic mixed model.}
    \label{fig:logistic_multi_sim_difference_dens_20231017}
\end{figure}

The synthetic datasets in this section are simulated according to the number of observations and parameter values detailed in Sect.~\ref{sec:logmm}. As with the previous section, we present plots comparing the R-VGAL and HMC posterior means in Figure~\ref{fig:logistic_multi_sim_means_20231017}, the ratio between the R-VGAL and HMC posterior standard deviations in Figure~\ref{fig:logistic_multi_sim_sds_20231017}, and the distributions of the differences between each method's posterior mean estimates and the true parameters in Figure~\ref{fig:logistic_multi_sim_difference_dens_20231017}.  The \Copy{multi_sims_logistic_results}{ 
R-VGAL and HMC posterior means are similar, though R-VGAL tends to slightly underestimate the random effect variance when compared to HMC.}

\clearpage
\subsection{Repeated simulations on the Poisson mixed model \label{sec:multi_sims_poisson}}
\begin{figure}
    \centering
    \includegraphics[width = \linewidth]{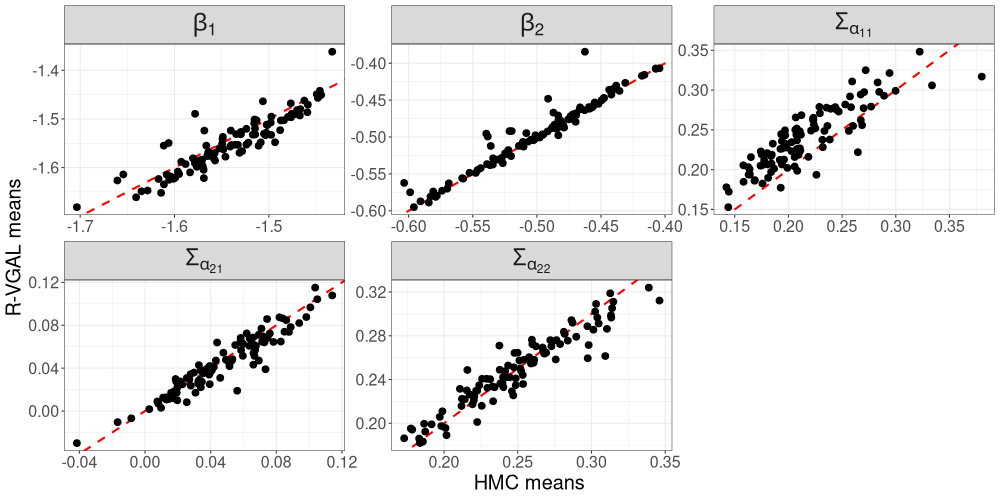}
    \caption{Plot of the R-VGAL and HMC posterior means for 100 datasets simulated from the Poisson mixed model.}
    \label{fig:poisson_multi_sim_means_20231224}
\end{figure}

\begin{figure}
    \centering
    \includegraphics[width = \linewidth]{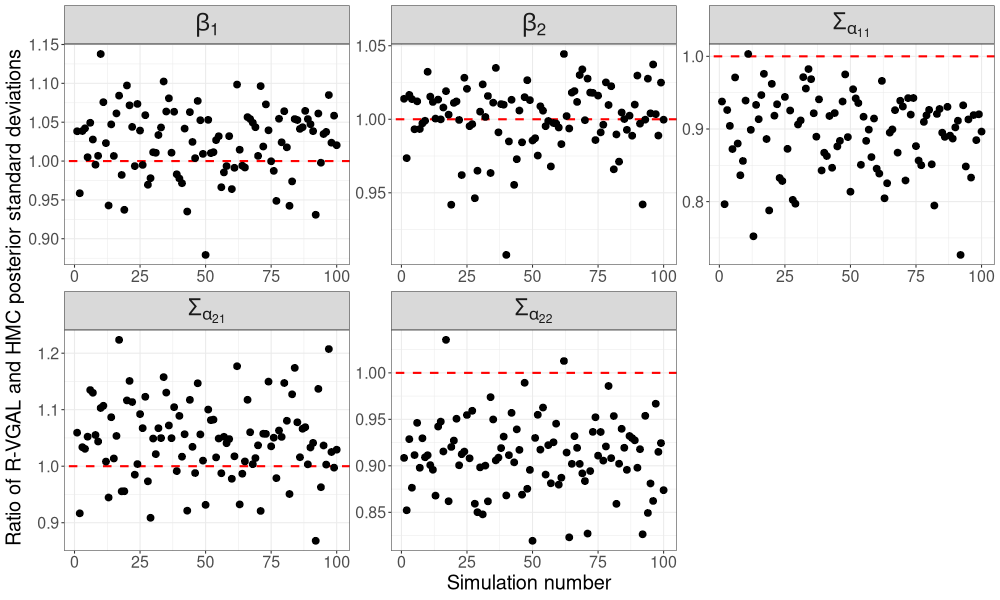}
    \caption{Plot of the ratio between the R-VGAL and HMC posterior standard deviations for 100 datasets simulated from the Poisson mixed model.}
    \label{fig:poisson_multi_sim_sds_20231224}
\end{figure}

\begin{figure}
    \centering
    \includegraphics[width = \linewidth]{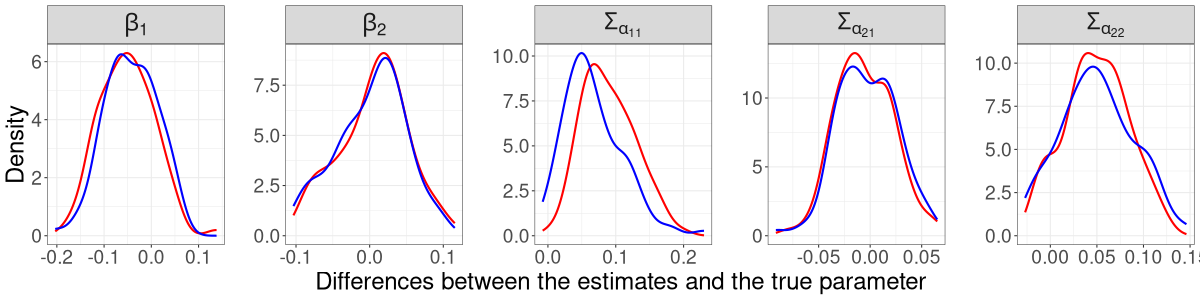}
    \caption{Comparison of the distribution of the differences between the R-VGAL posterior means and the true parameters (in red), against the distribution of the differences between the HMC posterior means and the true parameters (in blue), for 100 datasets simulated from the Poisson mixed model.}
    \label{fig:poisson_multi_sim_difference_dens_20231224}
\end{figure}

The synthetic datasets in this section are simulated according to the number of observations and parameter values detailed in Sect.~\ref{sec:poisson}. As with the previous sections, we present plots comparing the R-VGAL and HMC posterior means in Figure~\ref{fig:poisson_multi_sim_means_20231224}, the ratio between the R-VGAL and HMC posterior standard deviations in Figure~\ref{fig:poisson_multi_sim_sds_20231224}, and the distributions of the differences between each method's posterior mean estimates and the true parameters in Figure~\ref{fig:poisson_multi_sim_difference_dens_20231224}. \Copy{multi_sims_pois_results}{The R-VGAL and HMC posterior means and standard deviations are quite similar across simulations, although R-VGAL has a mild tendency for overestimating the parameter $\Sigma_{\alpha_{11}}$ and underestimating the posterior variance of $\Sigma_{\alpha_{11}}$ and $\Sigma_{\alpha_{22}}$ compared to HMC.} 

    \clearpage
\section{Convergence statistics for HMC}
\label{sec:hmc_convergence}

This section contains some statistics on the convergence of HMC in the simulated and real data examples in Sect.~\ref{sec:applications}. In particular, we show the effective sample size (ESS) and $\hat{R}$ values obtained from the \texttt{RStan} diagnostics. The ESS is a measure of the number of uncorrelated posterior samples, and the higher the ESS the better;
see ~\cite{gelman2003bayesian2} for further details on how STAN calculates ESS. The $\hat{R}$ statistic proposed by~\cite{gelman1992inference2} measures the ratio of the average variance of samples within each chain to the variance of the pooled samples across chains; an $\hat{R}$ close to 1 indicates that the chains have converged. \cite{gelman1992inference2} recommends that the independent Markov chains be initialized with diffuse starting values for the parameters and sampled until all values for $\hat{R}$ are below 1.1. 

The ESS and $\hat{R}$ values for each parameter in each of the examples we considered in Sect.~\ref{sec:applications} are shown in Table~\ref{tab:hmc_convergence}.

\begin{table}[ht]
        \centering
        \begin{tabular}{|c|c|c|c|}
            \hline
            Example & Parameter & ESS & $\hat{R}$ \\
            \hline
            Linear & $\beta_1$ & 42590.45 & 0.9999355\\
            Linear & $\beta_2$ & 46091.76 & 0.9999275\\
            Linear & $\beta_3$ & 45614.94 & 0.9999371\\
            Linear & $\beta_4$ & 44814.06 & 0.9999292\\
            Linear & $\sigma_\alpha$ & 39572.87 & 0.9999377\\
            Linear & $\sigma_\epsilon$ & 36754.35 & 0.9999034\\
            \hline
            Logistic & $\beta_1$ & 14116.662 & 1.0002822\\
            Logistic & $\beta_2$ & 12974.649 & 1.0000641 \\
            Logistic & $\beta_3$ & 26374.149 & 1.0001275 \\
            Logistic & $\beta_4$ & 31828.214 & 0.9999238 \\
            Logistic & $\tau$ & 3700.836 & 1.0004196 \\
            \hline
            Poisson & $\beta_1$ & 14313.12 & 1.0000291 \\
            Poisson & $\beta_2$ & 14433.41 & 1.0001494 \\
            Poisson & $\Sigma_{\alpha_{11}}$ & 20022.85 & 1.0000592 \\
            Poisson & $\Sigma_{\alpha_{21}}$ & 17313.42 & 1.0004479 \\
            Poisson & $\Sigma_{\alpha_{22}}$ & 19660.59 & 0.9999396 \\
            \hline
            Six City & $\beta_1$ & 2041.219  & 1.0001195\\
            Six City & $\beta_2$ & 21906.059 & 0.9999014 \\
            Six City & $\beta_3$ & 6902.207 & 0.9999044 \\
            Six City & $\tau$ & 1406.197 & 1.0007235 \\
            \hline
            Polypharmacy & $\beta_0$ & 8724.835 & 1.0001990 \\
            Polypharmacy & $\beta_{gender}$ & 7868.562   & 1.0003835 \\
            Polypharmacy & $\beta_{race}$ & 8810.641  & 1.0000568  \\
            Polypharmacy & $\beta_{age}$ & 17864.738  & 1.0000620 \\
            Polypharmacy & $\beta_{M1}$ & 17248.489  & 0.9999448  \\
            Polypharmacy & $\beta_{M2}$ & 15052.711  & 1.0000169  \\
            Polypharmacy & $\beta_{M3}$ & 14499.271  & 0.9999639  \\
            Polypharmacy & $\beta_{IM}$ & 31828.214 & 0.9999535  \\
            Polypharmacy & $\tau$ & 37289.351 & 1.0001344  \\
            \hline
        \end{tabular}
        \caption{Effective sample size and $\hat{R}$ values for the parameters in each model.}
        \label{tab:hmc_convergence}
    \end{table}
    \clearpage
\section{Additional examples \label{sec:add_examples}}
This section contains two additional examples: the first one applies the Poisson model in Sect.~\ref{sec:poisson} of the main paper to the Epilepsy dataset from~\cite{thallvail1990some}, and the second involves a bigger synthetic dataset simulated from the logistic mixed model in Sect.~\ref{sec:logmm} of the main paper.

\subsection{Real data example: Poisson mixed model \label{sec:epilepsy}}

In this example, we apply R-VGAL on the well-known Epilepsy dataset from~\cite{thallvail1990some}. This dataset includes $N = 59$ epileptic patients who were treated with either a new drug (Progabide) or placebo in a clinical trial. The response variable is the number of seizures
patients have during $n = 4$ follow-up periods. We index the patients as $i = 1, \dots, N$ and the responses for each patient as $j = 1, \dots, n$. We follow~\cite{tan2018gaussian} and use the following covariates: the logarithm of $1/4$ the number of baseline seizures (\texttt{Base}); the \texttt{Treatment}, coded as 1 for Progabide and 0 for placebo; the log-transformed and centred age, $\wtilde{\texttt{Age}}_i = \texttt{Age}_i - \frac{1}{N} \sum_{i=1}^N(\texttt{Age}_i)$, where $\texttt{Age}_i$ is the logarithm of the age of the $i$th individual; and the follow up period, \texttt{Visit}, coded as $\texttt{Visit} = -0.3$ for the first visit, $\texttt{Visit} = -0.1$ for the second, $\texttt{Visit} = 0.1$ for the third and $\texttt{Visit} = 0.3$ for the fourth. 

We consider the following model with random slope and random intercept:
\begin{align}
    y_{ij} &\sim \text{Poisson}(\lambda_{ij}), \\
    \log(\lambda_{ij}) &= \beta_0 + \beta_{base} \texttt{Base}_i + \beta_{treatment} \texttt{Treatment}_i + \beta_{age} \wtilde{\texttt{Age}}_i + \beta_{visit} \texttt{Visit}_{ij} + \alpha_{i,1} + \alpha_{i,2} \texttt{Visit}_{ij},
\end{align}
where $\balpha \equiv (\alpha_{i,1}, \alpha_{i,2})^\top \sim \Gau(\0, \bSigma_\alpha)$, with $\bSigma_\alpha = \L \L^\top$ and $\L = \begin{bmatrix}
        \exp(\zeta_{11}) & 0 \\
        \zeta_{21} & \exp(\zeta_{22})  
    \end{bmatrix}$.
In the algorithm, we consider the unconstrained parameters $\btheta = (\bbeta^\top, \bzeta^\top)^\top$, where $\bbeta \equiv (\beta_0, \beta_{base}, \beta_{treatment}, \beta_{age}, \beta_{visit})^\top$ and $\bzeta \equiv (\zeta_{11}, \zeta_{22}, \zeta_{21})^\top$. The gradient $\nabla_{\btheta} \log p(\y_i, \balpha_i \mid \btheta)$ and Hessian $\nabla^2_{\btheta} \log p(\y_i, \balpha_i \mid \btheta)$, which are necessary in the computation of the gradient and Hessian of the subject-specific log likelihood $\log p(\y_i \mid \btheta)$, are provided in Sect.~\ref{derivativepoissonregression} of the online supplement. 

The initial variational distribution we use is
\begin{equation*}
    p(\btheta) = q_0(\btheta) = 
    \Gau \left(
    \begin{bmatrix} \0 \\ \0 \end{bmatrix}, 
    \begin{bmatrix} \I_2 & \0_{2 \times 3} \\
    \0_{3 \times 2} & 0.1 \I_3
    \end{bmatrix}
    \right).
\end{equation*}
Using a $\Gau(0, 0.1)$ prior distribution for $\zeta_{11}$,  $\zeta_{22}$ and $\zeta_{21}$ leads to having 2.5th and 97.5th percentiles of (0.290, 3.485) for $\Sigma_{\alpha_{11}}$, (0.342, 3.577) for $\Sigma_{\alpha_{22}}$, and (-0.713, 0.713) for the off-diagonal entries $\Sigma_{\alpha_{21}}$ and $\Sigma_{\alpha_{12}}$. 

\begin{figure}
    \centering
    \includegraphics[width = 0.9\linewidth]{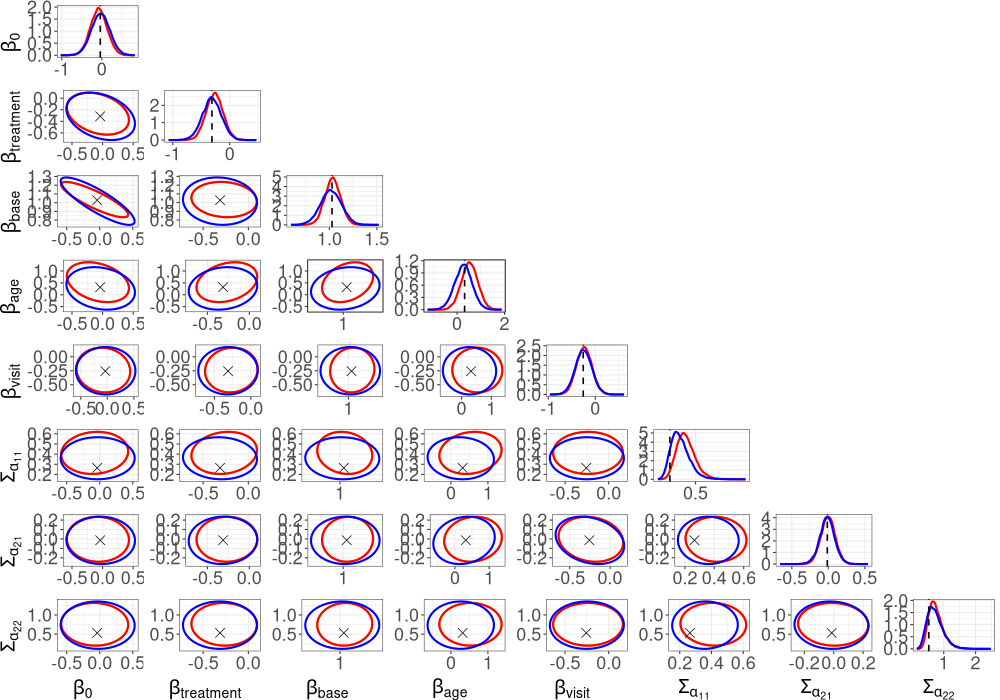}
    \caption{Exact posterior distributions from HMC (in blue) and approximate posterior distributions from R-VGAL with estimated gradients and Hessians (in red) for the Poisson mixed model experiment. Diagonal panels: Marginal posterior distributions with the maximum likelihood estimates marked using dotted lines. Off-diagonal panels: Bivariate posterior distributions with the maximum likelihood estimates marked using the symbol~$\times$.}
    \label{fig:epilepsy_posterior_temper10_S200_Sa200_20231018}
\end{figure}

As with the simulated Poisson model in Sect.~\ref{sec:poisson}, we run damped R-VGAL with $n_{damp} = 10$ observations and $K = 4$ steps per observation. We use $S_\alpha = 200$ Monte Carlo samples in the importance sampling step, and $S = 200$ samples to approximate the expectations with respect to $q_{i-1}(\btheta)$ in the R-VGAL updates of the variational mean and the precision matrix. Figure~\ref{fig:epilepsy_posterior_temper10_S200_Sa200_20231018} shows the marginal posterior distributions with maximum likelihood estimates of the parameters, along with bivariate posterior distributions as estimated using R-VGAL and HMC. We find that the estimates from R-VGAL and HMC agree quite well for all parameters, and posterior modes for all parameters from both methods are close to the maximum likelihood estimates.

\subsection{Big data example: Logistic mixed model \label{sec:big_data}}
For this experiment, we simulate data from the logistic mixed model in Sect.~\ref{sec:logmm} with $N = 5000$ and $n = 10$, resulting in a total of $50000$ observations. We find that it is necessary to increase the number of Monte Carlo samples to $S = 200$ and $S_\alpha = 200$ to achieve accurate posterior estimates in this example. The prior distribution and the settings for damping in R-VGAL are kept the same as in the example in Sect.~\ref{sec:logmm}.

\begin{figure}
    \centering
    \includegraphics[width = 0.9\linewidth]{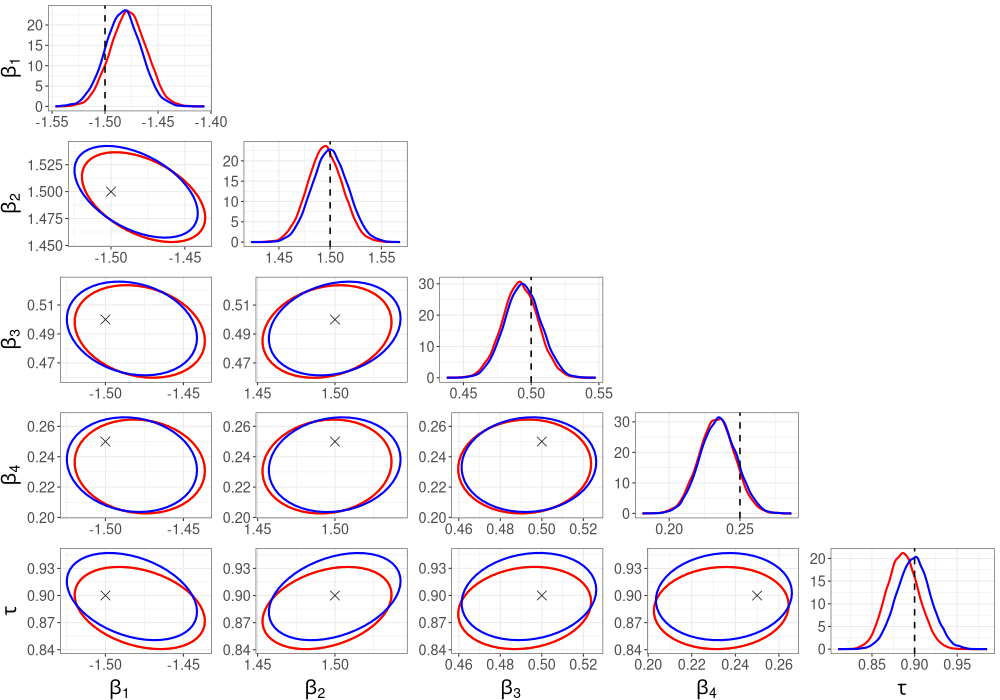}
    \caption{Exact posterior distributions from HMC (in blue) and approximate posterior distributions from R-VGAL with estimated gradients and Hessians (in red) for the logistic mixed model experiment with 50000 observations. Diagonal panels: Marginal posterior distributions with true parameters denoted using dotted lines. Off-diagonal panels: Bivariate posterior distributions with true parameters denoted using the symbol~$\times$.}
    \label{fig:logistic_posterior_temper10_S200_Sa200_20231201}
\end{figure}

Figure~\ref{fig:logistic_posterior_temper10_S200_Sa200_20231201} shows the marginal posterior densities, along with bivariate posterior plots from R-VGAL and HMC. As with previous simulations, the posterior distributions obtained using R-VGAL are very similar to those obtained using HMC.

    \section{Application of R-VGAL to models with other random effect structures \label{sec:crossed_nested}}

In the main paper, we considered the application of R-VGAL to models where the random effects are correlated within subjects (individuals), but independent between subjects. In practice, there are many cases where other random effect structures, such as crossed or nested random effects, are needed; see~\cite{pinheiro2006mixed, gelman2007data, west2014linear} or~\cite{papaspiliopoulos2023scalable} for examples. Here, we briefly discuss the application of R-VGAL to some classes of models with crossed or nested random effects. The implementation of R-VGAL to these models is left for future endeavours. 

Models with crossed effects are often used to model data that can be organised in the form of contingency tables between categorical variables. For example, consider a study of annual income, in which a number of characteristics from participants are recorded using categorical variables, such as their age range, sex, ethnicity, and highest level of education. In this case, the data can be organised into a multi-dimensional contingency table between age range, sex, ethnicity, and level of education. Participants who have the same combination of characteristics may have similar income levels, and the correlation between people in the same set of categories may be modelled with the addition of category-specific random effects. 

The notation we use in the following crossed effect model follows that in Chapter 11 of \cite{gelman2007data} and Sect. 2 of \cite{papaspiliopoulos2023scalable}. Suppose that there are $K$ categorical variables, and the $k$th variable has $L_k$ levels, for $k = 1, \dots, K$. The logarithm of the income of the $i$th individual may then be modelled as
\begin{equation}
    y_i = \x_i^\top \bbeta + \sum_{k=1}^K \alpha_{l_k[i]}^{(k)} + \epsilon_i, \quad \alpha_{l}^{(k)} \sim \Gau(0, \sigma_{\alpha}^2), \quad \epsilon_i \sim \Gau(0, \sigma_\epsilon^2), 
\end{equation}
for $i = 1, \dots, N$, where $\x_i$ denotes a vector of covariates associated with the fixed effects $\bbeta$, and $\alpha_l^{(k)}$ denotes the random effect associated with the $l$th level of the $k$th categorical variable. The notation $l_k[i]$ denotes the level of the $k$th category that the $i$th individual falls into; for example, if the first categorical variable in the model is age range, where the categories are 1 for 18--30 years old, 2 for 30--50 years old, and 3 for 50 years old and above, then $l_1[4] = 2$ means that the 4th individual in the dataset is in level 2 of the ``age" variable (between 30 and 50 years old). Here we have assumed that the random effects $\alpha_{l}^{(k)}$ have the same variance for all $l = 1, \dots, L$ and $k = 1, \dots, K$. Thus the parameters of interest in this model are $\btheta = (\bbeta^\top, \sigma_\alpha^2, \sigma_\epsilon^2)^\top$.

In this model, there are $G = L_1 \times L_2 \times \dots L_K$ combinations of levels. The R-VGAL updates would therefore be based on the log-likelihood of all individuals within the same combination (group). For each combination (group) $g = 1, \dots, G$, the vector of responses of people in this group is denoted as $\y_{g}$. Then for each group $g = 1, \dots, G$, the gradient of the group-specific log likelihood can be expressed via Fisher's identity as
\begin{equation}
\label{eq:crossed_grad_fisher}
    \nabla_{\btheta} \log p(\y_g \mid \btheta) = \int \nabla_{\btheta} \log p(\y_g, \alpha_{l_{1, g}}^{(1)}, \dots, \alpha_{l_{K, g}}^{(K)} \mid \btheta) \, p(\alpha_{l_{1, g}}^{(1)}, \dots, \alpha_{l_{K, g}}^{(K)} \mid \y_g, \btheta) \d \alpha_{l_{1, g}}^{(1)} \dots \d \alpha_{l_{K, g}}^{(K)},
\end{equation}
where, here, the notation $l_{k,g}$ denotes the level of the $k$th categorical variable associated with group $g$. The Hessian of the group log likelihood can be similarly expressed via Louis' identity~\eqref{eq:louis_identity}, which we do not restate here. The gradient and Hessian of the group log likelihood can then be approximated using the importance-sampling-based approach described in Sects.~\ref{sec:fishers_identity} and~\ref{sec:louis_identity}. Note that there are analytical formulae for the gradient (and Hessian) in this case, as the model is linear; but this approach is applicable to a wide class of GLMMs with crossed random effects.

R-VGAL is also applicable to models with nested random effects. One popular example of such models is that of students being nested within classes, which are then nested within schools (see, for example, Chapter 4 of~\cite{west2014linear}). Suppose that we are interested in the final exam marks of Year 12 students across a number of schools in the year 2023. If we write the final mark of the $k$th student in the $j$th class at the $i$th school as $y_{ijk}$, for $i = 1, \dots, N$, $j = 1, \dots, n_i$ and $k = 1, \dots, n_{ij}$, then a simple linear model with one random intercept at both the school level and the class level is
\begin{equation}
\label{eq:student_level}
    y_{ijk} = \x_{ijk}^\top \bbeta + \alpha_i + \gamma_{ij} + \epsilon_{ijk}, \quad \alpha_i \sim \Gau(0, \sigma_\alpha^2), \quad \gamma_{ij} \sim \Gau(0, \sigma_\gamma^2), \quad \epsilon_{ijk} \sim \Gau(0, \sigma_\epsilon^2),
\end{equation}
where $\x_{ijk}$ is a vector of fixed covariates for the $k$th student (which may include, for instance, the average number of hours they spend studying per week, or the average number of hours slept per night), $\bbeta$ are the corresponding fixed effects, $\alpha_i$ is the random effect associated with the school that the student attends, $\gamma_{ij}$ is the random effect associated with the class they are in at their school, and $\epsilon_{ijk}$ is an error term associated with each student.  

At the class level, the model~\eqref{eq:student_level} may be written as
\begin{equation}
\label{eq:class_level}
    \y_{ij} = \X_{ij} \bbeta + \1_{n_{ij}} \alpha_i + \1_{n_{ij}} \gamma_{ij} + \bepsilon_{ij}, \quad \alpha_i \sim \Gau(0, \sigma_\alpha^2), \quad \gamma_{ij} \sim \Gau(0, \sigma_\gamma^2), \quad \bepsilon_{ij} \sim \Gau(\0, \bSigma_{\epsilon, i}), 
\end{equation}
where $\y_{ij} \equiv (y_{ij1}, \dots, y_{ijn_{ij}})^\top$ is a vector containing final exam marks for all students in class $j$ of school $i$, $\X_i \equiv (\x_{ij1}^\top, \dots, \x_{ijn_{ij}}^\top)^\top$ is a design matrix containing fixed covariates from all students in class $j$ of school $i$, $\1_{d}$ denotes a $d \times 1$ vector of ones, and $\bepsilon_{ij} \equiv (\epsilon_{ij1}, \dots, \epsilon_{ijn_{ij}})^\top$, for classes $j = 1, \dots, n_i$ and schools $i = 1, \dots, N$. For simplicity, we assume here that the error terms $\epsilon_{ijk}, k = 1, \dots, n_{ij}$, are uncorrelated, so that $\bSigma_{\epsilon, i} = \sigma_\epsilon^2 \I_{n_{class, i}}$, where $\I_d$ denotes a $d \times d$ identity matrix, and $n_{class, i}$ is the number of students in each class at school $i$, which we assume does not change with class; that is, we let $n_{ij} = n_{class,i}$, for $j = 1,...,n_i$ and $i = 1,...,N$.

We will now go one step further and write the model at the school level as
\begin{equation}
\label{eq:school_level}
    \y_{i} = \X_{i} \bbeta + \Z_i \alpha_i + \W_{i} \bgamma_{i} + \bepsilon_{i}, \quad \alpha_i \sim \Gau(0, \sigma_\alpha^2), \quad \bgamma_{i} \sim \Gau(\0, \bSigma_{\gamma_i}), \quad \bepsilon_{i} \sim \Gau(\0, \I_{n_{i}} \otimes \bSigma_{\epsilon, i}), 
\end{equation}
where $\otimes$ denotes the Kronecker product between two matrices, $\X_i \equiv (\X_{i1}, \dots, \X_{in_i})^\top$, $\Z_i \equiv \1_{n_i n_{class, i}}$,
${\W_i = \I_{n_i}} \otimes \1_{n_{class,i}}$, 
$\bgamma_i \equiv (\gamma_{i1}, \dots, \gamma_{in_i})^\top$ is a vector of all random effects from classes $j = 1, \dots, n_i$ in school $i$, and $\bepsilon_i \equiv (\bepsilon_{i1}^\top, \dots, \bepsilon_{in_i}^\top)^\top$ is a vector of random error terms for all students in school~$i$, $i = 1, \dots, N$. Here we allow the random effects $\gamma_{ij}, j = 1, \dots, n_i$ to be correlated between classes in the same school, so that for each school $i = 1, \dots, N$, the covariance matrix $\bSigma_{\gamma_i}$ is potentially dense. We also assume that the school-specific random effects $\alpha_i$ are independent and identically distributed for all schools $i = 1, \dots, N$.

The parameters of interest in this model are $\btheta = \{\bbeta, \sigma_\alpha^2, \bSigma_{\gamma_1}, \dots, \bSigma_{\gamma_N}, \sigma_\epsilon^2\}$. To make inference on these parameters, the R-VGAL updates can be performed by processing the data one school at a time. That is, for $i = 1, \dots, N$, the R-VGAL updates are made in terms of the gradient and Hessian of the school-specific log-likelihood, $\nabla_{\btheta} \log p(\y_i \mid \btheta)$ and $\nabla^2_\theta \log p(\y_i \mid \btheta)$. Using Fisher's identity~\eqref{eq:fishers_identity}, the gradient $\nabla_{\btheta} \log p(\y_i \mid \btheta)$ can be written as
\begin{equation}
\label{eq:nested_grad_fisher}
    \nabla_{\btheta} \log p(\y_i \mid \btheta) = \int p(\alpha_i, \bgamma_i \mid \btheta) \nabla_{\btheta} \log p(\y_i, \alpha_i, \bgamma_i \mid \btheta) \d \alpha_i \d \bgamma_i, \quad i = 1, \dots, N,
\end{equation}
where the term $\log p(\y_i, \alpha_i, \bgamma_i \mid \btheta)$ can be expanded further as
\begin{equation}
    \log p(\y_i, \alpha_i, \bgamma_i \mid \btheta) = \log p(\y_i \mid \alpha_i, \bgamma_i, \btheta) + \log p(\alpha_i, \bgamma_i \mid \btheta),
\end{equation}
with
\begin{align*}
    p(\y_i \mid \alpha_i, \bgamma_i, \btheta) &= \Gau(\X_{i} \bbeta + \Z_i \alpha_i + \W_{i} \bgamma_{i}, \bSigma_{\epsilon, i} \otimes \I_{n_i}), \\
    p(\alpha_i, \bgamma_i \mid \btheta) &= 
    \Gau \left(
    \begin{bmatrix}
        0 \\ \0 
    \end{bmatrix}, 
    \begin{bmatrix}
        \sigma_\alpha^2 & \0^\top \\ 
        \0 & \bSigma_{\gamma_i}
    \end{bmatrix} 
    \right), \quad i = 1, \dots, N,
\end{align*}
based on~\eqref{eq:school_level}. Approximation of the gradient $\nabla_{\btheta} \log p(\y_i \mid \btheta)$ can then proceed via the importance-sampling-based approach in Sect.~\ref{sec:fishers_identity}, and a similar procedure for the Hessian $\nabla_{\btheta} \log p(\y_i \mid \btheta)$ can be done as shown in Sect.~\ref{sec:louis_identity}. As the model is linear in this case, we note that the log-likelihood, gradient, and Hessian are available analytically. However, this approach is applicable to a wide class of GLMMs with nested random effects.




\end{document}